%% file: sample1.tex
\documentclass[11pt]{book}
\usepackage{Wiley-AuthoringTemplate}
\usepackage{chapterbib}

\usepackage{subfigure}
\DeclareMathOperator*{\argmin}{arg\,min}

\usepackage{amsfonts}

\usepackage[sectionbib,authoryear]{natbib}% for name-date citation comment the below line
\makeindex% for index generation

\begin{document}

\frontmatter

\mainmatter

\include{ch01}

\include{ch02} 
\include{ch03} 
\include{ch04}
\include{ch05}

\backmatter
\include{app01}
\include{app02}

\printindex

\end{document}

%% file: ch01.tex
\chapter[This is Chapter One Title containing authors and affiliations]{Cyber Insurance for Cyber Resilience\protect}%\footnote{This is a sample for chapter footnote.}}

%\author*[1]{Shutian Liu}
\author[1]{Shutian Liu}
\author[1]{Quanyan Zhu}

\address[1]{\orgdiv{Tandon School of Engineering}, 
\orgname{New York University}, 
\postcode{11201}, \countrypart{NY}, 
     \city{Brooklyn}, \street{370 Jay Street}, \country{US}}%

%\address[2]{\orgdiv{II Author Organization Division Name}, 
%\orgname{Organization Name}, 
%\postcode{Postal Code}, \countrypart{Part of the Country}, 
%     \city{City Name}, \street{Street Name}, \country{Country}}%

%\address[3]{\orgdiv{II Author Organization Division Name}, 
%\orgname{Organization Name}, 
%\postcode{Postal Code}, \countrypart{Part of the Country}, 
%     \city{City Name}, \street{Street Name}, \country{Country}}%

\address*{Corresponding Author: Shutian Liu; \email{sl6803@nyu.edu}}

\maketitle% This tag is required to print author and address in the output

\begin{abstract}{}
Cyber insurance is a complementary mechanism to further reduce the financial impact on the systems after their effort in defending against cyber attacks and implementing resilience mechanism to maintain the system-level operator even though the attacker is already in the system.
This chapter presents a review of the quantitative cyber insurance design framework that takes into account the incentives as well as the perceptual aspects of multiple parties. 
The design framework builds on the correlation between state-of-the-art attacker vectors and defense mechanisms.
In particular, we propose the notion of residual risks to characterize the goal of cyber insurance design.
By elaborating the insurer's observations necessary for the modeling of the cyber insurance contract, we make comparison between the design strategies of the insurer under  scenarios with different monitoring rules. 
These distinct but practical scenarios give rise to the concept of the intensity of the moral hazard issue.
Using the modern techniques in quantifying the risk preferences of individuals, we link the economic impacts of perception manipulation with moral hazard.
With the joint design of cyber insurance design and risk perceptions, cyber resilience can be enhanced under mild assumptions on the monitoring of insurees' actions.
Finally, we discuss possible extensions on the cyber insurance design framework to more sophisticated settings and the regulations to strengthen the cyber insurance markets.

\end{abstract}

%\begin{abstract}{}
%Abstract text. Abstract text. Abstract text. Abstract text. 
%\end{abstract}


\section{Introduction}
\label{sec:intro}
Cyber resilience refers to the abilities of a cyber system or a cyber physical system to be resistant to potential harms, be repairable from cyber attacks and system failures, and be sustainable after a pause of services or a shut-down of networked communications.
Cyber resilience extends the notion of cyber security, for resiliency counts for post-damage restorations while aims to secure systems are set up at the ex-ante stage to defend against interim harms.
Possible ways to enhance the level of cyber resilience includes setting up anti-virus software and firewall, deploying intrusion detection systems (IDSs) and moving target defense systems, and using autonomous defense.
These traditional techniques build on increasing system security against possible attack vectors including phishing, cryptojacking, and advanced persistent threats (APTs). 
However, not all attacks can be successfully defended, and many system failures are unavoidable.

Cyber insurance is a unique approach to enhance cyber resilience among the other traditional defense mechanisms.
It is a complementary mechanism to further reduce the financial impact on the systems (e.g., critical infrastructure operators, autonomous system users, and enterprise network owners) after their efforts in defending against cyber attacks and implementing resilience mechanisms to maintain the system-level operation despite the fact that the attacker is already in the system. The perfect defense and resilience mechanisms that can prevent or deter the attacker from inflicting significant damages are either cost-prohibitive or impossible to attain without compromising the usability. There exist residual cyber risks even though the system has deployed sophisticated defense mechanisms. Cyber insurance is one last resort to mitigate the damage on individual systems through financial means so that business or operational continuity of the system can be ensured and further recovery is economically feasible. It is particularly important to individuals or small or midsize businesses who do not have sufficient cybersecurity expertise and cannot afford a significant investment in cybersecurity for their systems.

\begin{figure}
\centering
\includegraphics[width=0.8\textwidth]{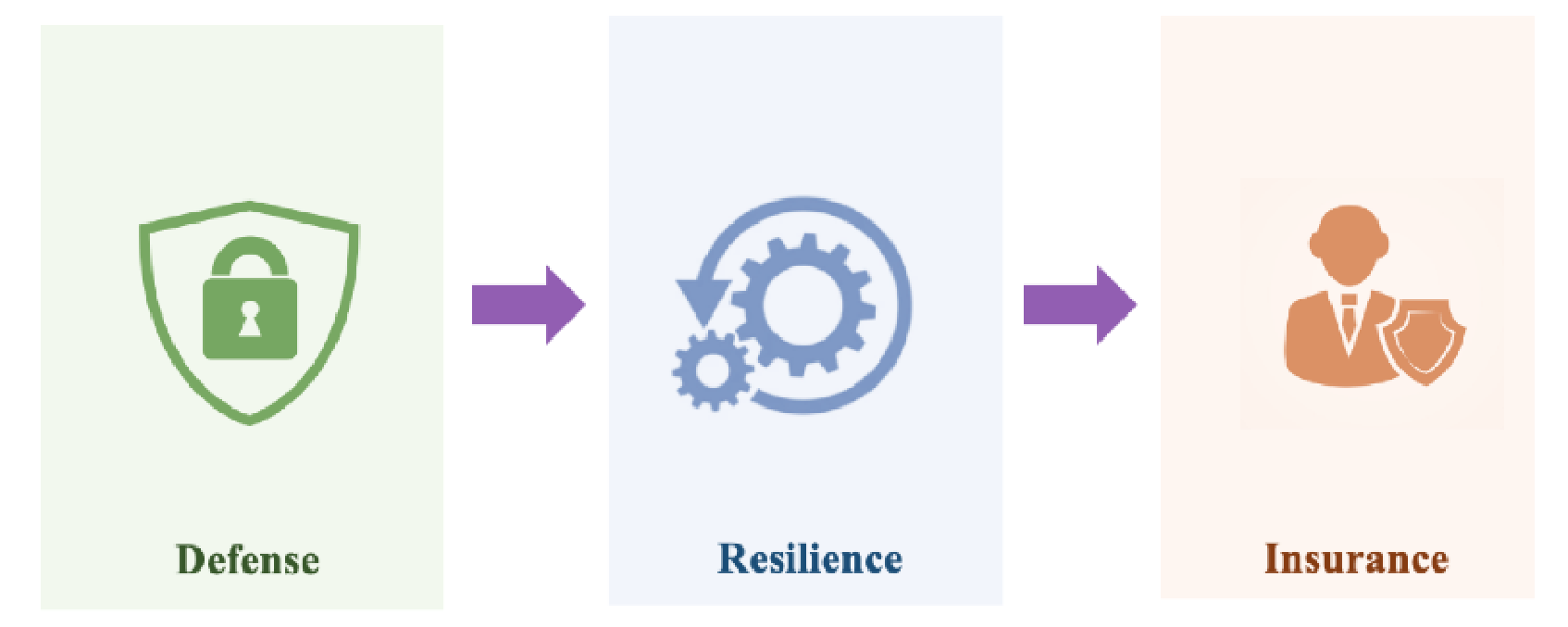}
\caption{Insurance is a complementary mechanism to provide a socio-economic layer of resilience to the system protected under a technical security and resilience mechanism.
\label{fig:insurance}}
\end{figure}

From the user's perspective, cyber insurance is not a direct securtiy investments.
It is an agreement initiated by payments from the users to the insurer and followed by the insurer's promises of covering the costs of cyber losses to be encountered by the users.
Cyber insurance functions through risk sharing.
Risk sharing exists between the users and the insurer, as it can be observed from how a cyber insurance contract is established. 
A prominent use case of cyber insurance where the users benefit from risk sharing is ransomware which has recently spread and affected many victims including large corporations (e.g., Colonial Pipeline ransomware attack \citep{hobbs2021colonial} and Honda ransomware attack \citep{chen2017automated}) and many small businesses, including local high schools and hospitals. Attackers are financially motivated to gain by cryptoviral extortion. 
The victims would face the to-pay-or-not-to-pay dilemma. Without paying the ransom, the services and the businesses supported by the system will be disrupted. However, by paying the ransom, the victims will suffer financially and irrevocable damage to the system and their reputation. Cyber insurance is a risk-sharing mechanism that can transfer the residual risks to a third party (the insurers) to mitigate the loss due to the ransom payment as well as the disruptions caused by the attack.
Besides, there is also risk sharing among the users, for the existence of the cyber insurance markets depends on the population of the insured users.

Cyber insurance adds to the users' strategy pools for defending their systems, since users can invest on system security and purchase cyber insurance at the same time. 
While the security investments can stochastically reduce the cyber risks, contacting on cyber insurance protects the users against potential severe losses from low-probability yet fatal attacks or failures. For mid and small scale businesses and companies, cyber insurance may be the only feasible way to defend their systems. The reason lies in that, with the advances of the attack vectors, there is a need to involve experienced and sophisticated cybersecurity expertise or professional teams to build protection systems or to form security consulting departments. These approaches may be too costly for mid to small scale business owners.
Cyber insurance comes more handy since the expensive procedure of hiring and training of individuals is taken care of by the insurance company. 
What the owners need to do is to choose a reliable third-party and contract with it to obtain a reasonable coverage of future cyber losses.
Cyber insurance creates a win-win situation for the insurers and the users. 
A successful cyber insurance contract not only benefits the users by reducing their financial losses, but also produce profits to the insurance companies.
Therefore, it is beneficial to create cyber insurance markets to enable an ecosystem in which system users and operators can further mitigate their cyber risks.

To this end, there is a need to understand the challenges in the ecosystem that prohibit the implementation of cyber insurance.
These challenges arise because of the influences on the ecosystem, which include the technologies for understanding cyber losses to enable insurance plans, the agents (customers) in the cyber insurance, and the economic relations between the insurer and the agents that affect the incentives for purchasing the insurances and the agents' behaviors when they are insured.
From the aspect of the technologies, the coverage plan of the insurance is challenging to determine because of the difficulties in cyber risk assessment.
The reason is two-fold.
Firstly, due to the existence of different types of cyber attacks (such as human-layer attacks caused by employees, cyber layer attacks targeting the IT, and physical layer attacks), it is impossible for the network users to locate the actual cause of cyber risks. 
In other words, the origin of cyber risks are mixed, as attacks from different layers are interdependent.
Secondly, the insurer has no incentive to cover the cyber threats that are likely to be successfully defended by common defense mechanisms.
Therefore, it is crucial for the insurer to figure out the feasible cyber defense actions of their customers and design coverage plans keyed to those defenses that are not included.
From the aspect of human perceptions, risk preferences play an essential role in establishing insurance contracts.   
On the one hand, the users are often risk-averse, since they wish to mitigate the uncertainty they are facing and aim to avoid severe fatal outcomes despite of their low possibilities.
On the other hand, the insurer needs to be risk neutral, since their assessment of the risk should be rational in order to provide revenue-maximizing risk-sharing insurance services.
Due to the challenges of risk assessments caused by the aspect of technologies, the risk preferences of the users and the insurer add an additional dimension of difficulty in the quantification of the perceived cyber losses.
From a practical view point, elicitation techniques can aid the calibration of human risk preferences. 
However, there is no universal agreement on a specific mathematical description for representing a risk attitude.
Moreover, cyber risks are more challenging to evaluate since immaterial resources in the Informational-Technology (IT) systems such as data and information are often involved.
Hence, choosing a proper and convenient model to assess the risks is still an obstacle sitting in front of the implementation of cyber insurances.
From an economic perspective, insurance design is challenging since dilemmas may occur when the insurer and the insured agents are asymmetric in terms of knowledge about the environment and information of the actions.
The adverse selection situation and the moral hazard issue, among the others, are the most common phenomena.
While the former appears due to information asymmetry before establishing an insurance contract, the latter occurs when the insurer cannot observe the real action taken by the insurers after the contract has been signed.
They both create challenges for designing and implementing cyber insurances, since the insurance contract may be be profitable anymore for the insurer in presence of these dilemmas.

\begin{figure}
\centering
\includegraphics[width=0.8\textwidth]{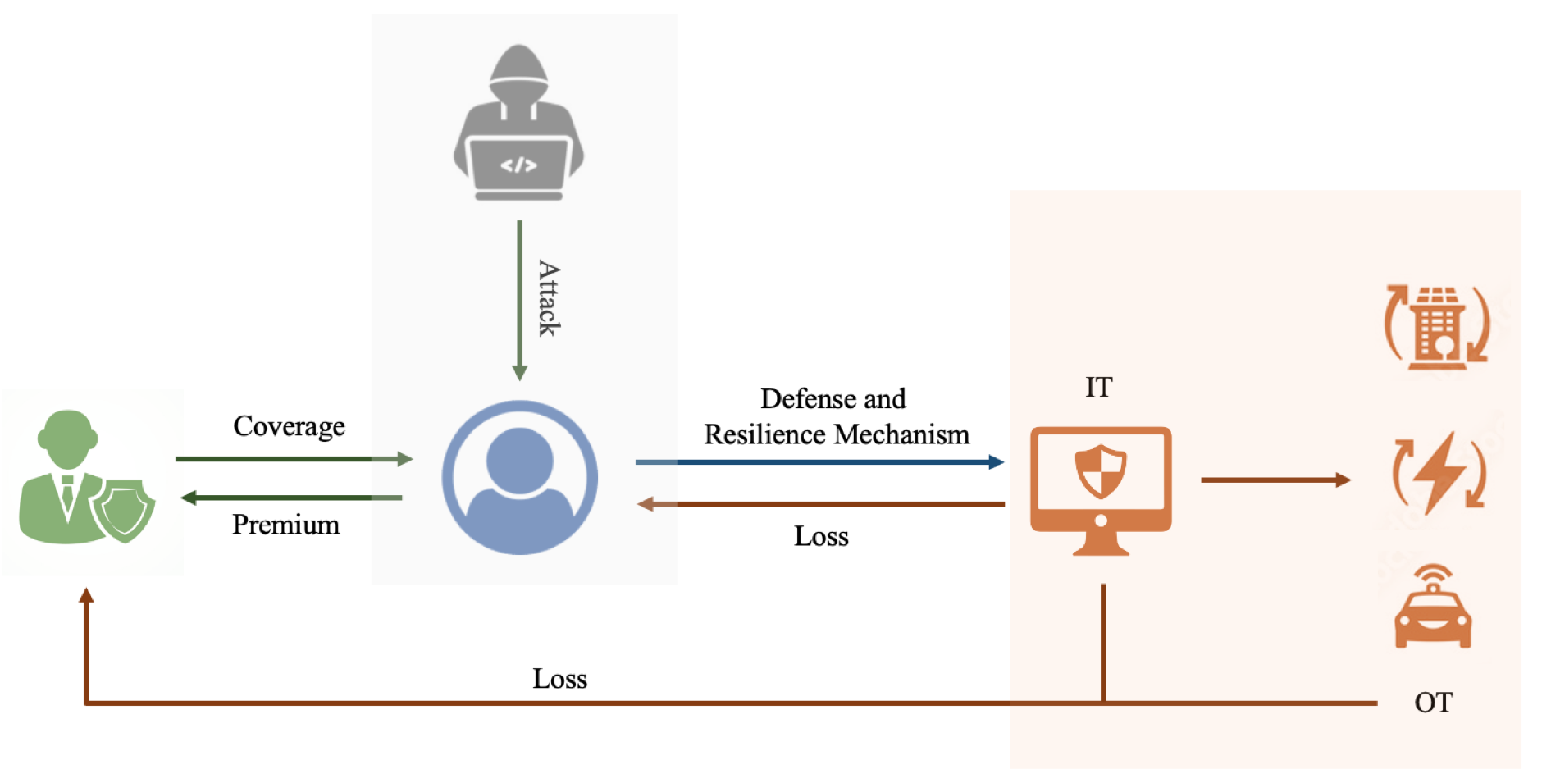}
\caption{The cyber insurer determines the premium and the coverage for a user who operates IT/OT systems and interacts with an attacker. 
\label{fig:framework}}
\end{figure}

This chapters presents a review of the quantitative cyber insurance design framework and discusses about how the challenges mentioned previously can be overcome with the help of modern techniques.
Fig. \ref{fig:framework} illustrates the framework by depicting the interactions among the three players, the insurer, the insuree, and the attacker. The insuree defends its IT system and protects its Operational Technology (OT) systems (e.g., smart buildings, smart grids, and autonomous vehicles) from damages. The attacker can exploit vulnerabilities in the IT system to take over the OT resources. The insurer determines the level of coverage as well as the premium to reduce the total risk in the system and maximize his revenue.

We explicitly introduce several popular and state-of-the-art attack models with the corresponding insured targets in Section \ref{sec:attack models}.
We discuss the potential defense mechanisms in accordance with the attack models in Section \ref{sec:defense mechanisms}.
The quantitative cyber insurance design framework will be introduced in Section \ref{sec:insurer observations} after we elaborate the insurer's observations necessary for the modeling of the cyber insurance contracts.
The review of risk preference modeling and the relation between risk preferences and system security in cyber insurance contracts in included in Section \ref{sec:risk modeling}.
In Section \ref{sec:preference manipulation}, we discuss a novel situation where human perceptions can be shaped using information. The economic impact of this perception manipulation is linked with the moral hazard issue.
Possible extensions of the cyber insurance design framework to the dynamic setting with be discussed in Section \ref{sec:dynamic}.
%We will consider OT into the design framework in Section \ref{sec:cps insurance} to discuss cyber insurance for cyber-physical systems (CPSs).
Regulations on cyber insurance will appear Section \ref{sec:regulations}, including accountability and compliance issues.

\section{Attack Models and Insured Targets}
\label{sec:attack models}
The user can also be referred to as the defender of a system.
The defender aims to protect her system against potential harms from the attacker, who acts strategically and stealthily to achieve his attack objectives. The goal of the attacker can be categorized using CIA triad (i.e., confidentiality, integrity, and availability) as shown in Fig. \ref{fig:triad}. For example, ransomware attackers often aim to compromise the availability of the data for critical services, and they choose the amount of ransom depending on the wealth and the willingness-to-pay of the victims. The Target data breach is another attack that aims at data confidentiality. The attacker gained access to Target servers through the credentials of a third-party vendor in 2013. It has been reported that Target claimed a total loss of $290$ million as a result.

\begin{figure}
\centering
\includegraphics[width=0.6\textwidth]{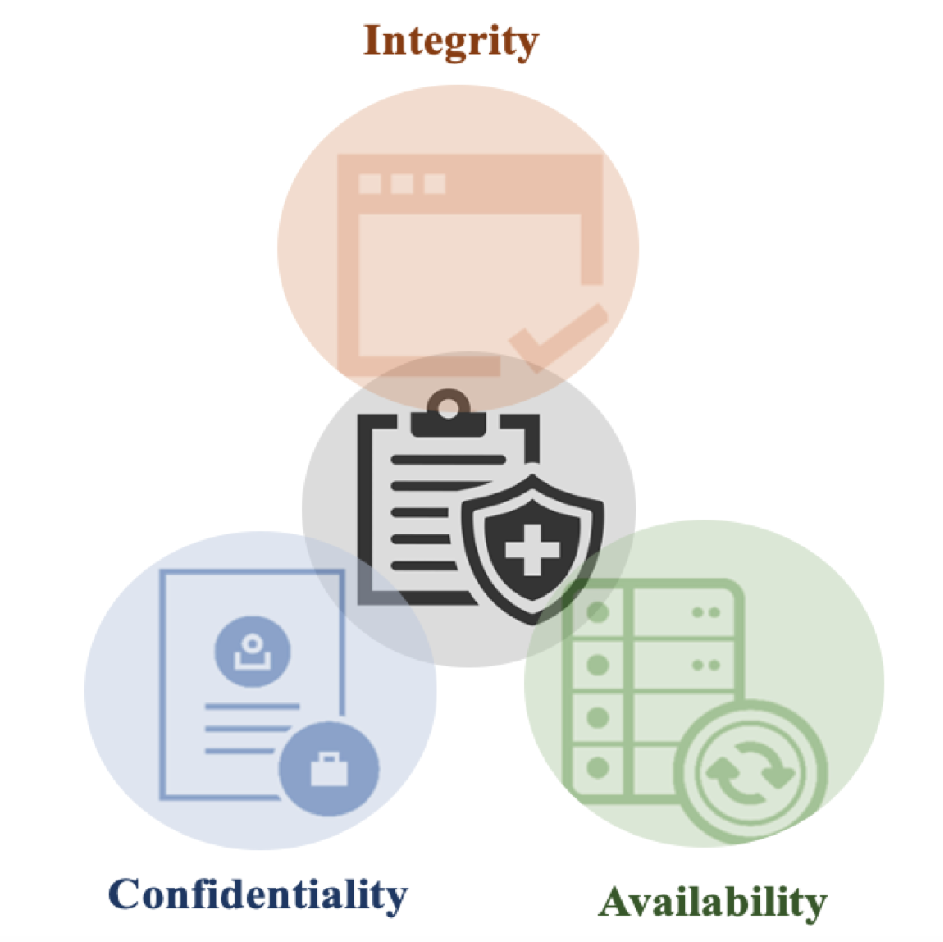}
\caption{The insurance can cover multiple types of attacks that aim to compromise different aspects of the system, including integrity, confidentiality, and availability.
\label{fig:triad}}
\end{figure}

This section introduces several types of popular and state-of-the-art attack models and the corresponding targets to insure. 
We divide the discussion into three parts, namely, the human-layer attacks caused by or through insiders, the cyber-layer attacks that target the IT, and the physical-layer attacks which can harm the physical device and equipment.
We not only discuss cyber threats which mainly fall into one of the three layers, but also, and more importantly, mention the interdependencies among the layers.
For example, an attack vector targeting the cyber layer can be initiated using a phishing email, meaning that this attack can be successfully defended if the employees are well-trained.
The connections among the layers make it challenging for the insurer to conduct risk assessments.
They also complicate the deployment of cyber defense mechanisms and raise the challenges of determining the coverage plan of the insurance.

\subsection{Human-layer Attacks}
The human-layer attack is a special type of attack where the attacks are not directly aiming at the cyber systems. 
On the contrary, human-layer attacks are those attacks that influence the cyber systems through the actions of the targeted individuals.
Phishing attack is one of the most common example, since the phishing emails and text messages aim to utilize the lacks of awareness of individuals to either steal private information or harm the networked system through a trusted user.
We focus on the insider threats \citep{liu2018detecting,hunker2011insiders}, where the attacks are performed by the so-called insiders, that is, individuals who are advantageous or privileged in accessing certain systems.
The notion of insider is fairly general.
An employee of a company or a business can be referred to as an insider. 
A software development engineer who has knowledge about the coding of an IT system can create insider threats.
Any individual who has physical access to the computer terminals can also be considered as insiders.

Insider threats, compared to other cyber losses caused by an arbitrary individual can be more severe and sometimes fatal to the IT systems.
The reason lies in but not limited to the following three points.
Firstly, insiders may have extra knowledge about the weaknesses of an IT system compared to other people.
This advantage can be utilized to directly target the attacks at the system vulnerabilities, causing massive losses to the IT systems.
Secondly, insiders can act more stealthily, since they are classified as authorized or trusted users.
The behavior of an employee accessing documents and resources from the intranet is common.
Hence, it is more challenging to detect it as a malicious behavior when an authorized individual manipulates or collects certain data.
Thirdly, insider threats can happen at anytime due to the advantages in accesses.
Unless the IT system is constantly monitored, it is difficult to identify malicious behaviors.
However, monitoring and verifying all behaviors happening at all time is costly and may not be a feasible approach for many businesses and companies.

Several taxonomies of insiders and insider threats have been discussed in the literature \citep{bishop2009case,pfleeger2009insiders}.
According to the survey \citep{liu2018detecting}, one taxonomy partitions insiders into masqueraders, traitors, and unintentional perpetrators. 
While traitors and unintentional perpetrators mainly create insider threats including data exfiltration, violation against data integrity or availability, and sabotage of Information and Communication Technology (ICT) systems \citep{silowash2012common}, masqueraders contributes to a richer domain of insider threats.
The reason lies in that the kill chain of a masquerader involves several phases, where difference threats can be posed \citep{hutchins2011intelligence}.
Examples of threats posed by masqueraders include network and database vulnerability scans, email spam and phishing, and distributed denial-of-service (DDoS) \citep{salem2008survey}.

On the one hand, since the insiders are already in the system, the above threats cannot be avoided by setting up firewalls or anti-virus software.
On the other hand, businesses and companies can turn to cyber insurance to mitigate the financial risk of these insider threats.
Although cyber insurance is a feasible solution to insider threats, the targets to insure needs further investigation.
Indeed, the insured targets in this scenario can vary from individual computers in the ICT system to datasets that contain essential information.
The reason lies in that insiders may adopt any attack vector to achieve his goals. 
The threats caused by APTs \cite{salem2008survey} can target arbitrary entry point of a networked system.

\subsection{Cyber Layer Attacks}
There are many attack vectors which can be classified as directly targeting the IT systems. 
For example, cryptojacking, viruses and worms, and payment frauds.
We focus on two types of attacks, namely APTs and data breaches, as the nominal ones to represent the state-of-the-art strategies from the attackers.
Both of them can lead to fatal consequences to businesses and companies.
The APTs have been proved to be one of the most dangerous attacks, since the hackers are directly targeting the systems or the business. They will not give up before they manage to achieve their malicious goals.
Data breaches often occur in a stealthy way that the defenders will not be aware of the stolen or manipulated data until they have somehow realized that their business secrets have been taken advantage of by their opponents in the market.

\subsubsection{Advanced Persistent Threats}
APTs are the most dangerous cyber threats which can involve various state-of-the-art attack methods and can last for a sufficiently long period until the attackers achieve their goals \citep{alshamrani2019survey}.
Perhaps the most infamous example of an APT is the Stuxnet virus \citep{farwell2011stuxnet}.
Initially infected through the universal-serial-bus (USB) devices in June 2010, the Stuxnet computer worm eventually caused damage to the control systems
of a nuclear-enrichment plant in Iran.
Due to the effectiveness of Stuxnet, many variances of Stuxnet started to emerge, including Duqu, Shamoon, Triton, etc \citep{al2018stuxnet}.
These APTs are challenging to detect or notice, for they often involve a lateral movement phase in their attacks.
Hence, it is insufficient to merely rely on defense systems to avoid being infected by APTs.
Furthermore, since APTs often have specific attack targets, they can cause the most deadly damages to a system.
Therefore, businesses and companies should be prepared, to a reasonable level, for recovering from potential attacks in the form of APTs.

Insuring against APTs is generally a challenging task due to the fact that APTs can be sophisticated enough to stealthily target the most essential components in a system.
Nevertheless, feasible insuring targets includes, but not limited to, the following.
The dataset containing sensitive information is a reasonable target, since the market strategies or business plans of companies are determined based on it.
It is also necessary to insure the main communication links in an ICT system, for the interconnections between sub-networks and information flows for maintaining synchronous functions rely heavily on these links.
From the example of the Stuxnet attack, we know that critical infrastructures can also be the target of APTs.
Therefore, the functionality of these physical components in the OT system should also be guaranteed.
Companies are highly recommended to purchase cyber insurance for their base stations or 
networked computer systems to obtain compensations under unavoidable APTs.

\subsubsection{Data Breaches}
Data breaches can happen everyday at any place \citep{liu2018understanding,barona2017survey}.
Confidentiality is the main concern regarding stolen data or missing information.
The data stolen can contain private information, including personal contract information, identification information, and even individual medical history.
These information can not only be used in marketing and advertising, but also be taken advantage of by malicious purposes.
Hence, to avoid severe consequences caused by data breaches, there is a need to understand the value of the stolen data and how the data can be utilized.
However, it is challenging to do so in practice.

There can be two possible scenarios of data breaches.
In the first scenario, data is stolen or missing and the users are aware of it. 
Hence, the user can quantify the amount of stolen data and verify its usage in the system.
Then, the user can estimate the potential loss based on the performance degradation without the stolen data.
In other words, since the users have knowledge of what data is stolen, they can approximately determine the corresponding losses to their businesses because of the data breaches.
In the second scenario, data is duplicated by malicious attackers.
In this case, the users still hold the data duplicated by the attacker.
Hence, it is more challenging for the user to verify what data has been leaked. 
Accordingly, it is more difficult to quantify the loss because of the leakage. The only way to infer the loss is to compare the system performance with potential competitors in the market.
Take an advertisement company as an example.
When data breach take place, the companies dataset containing customer preferences can be duplicated by another advertisement company. 
The company being attacked is not aware of the data breach. 
But advertisement designers from the company can conjecture about the data breach if they observe advertisements which target at
their potential customers.

The insured targets for data breaches should be distinguished according to the two scenarios discussed previously.
For the first scenario, since the value of the stolen data can be quantified by the users, the insured target can be simply the data.
For the second scenario, since the users can only be aware of the duplication from exogenous responses, quantification of the value of the duplicated data is challenging.
Therefore, an alternative insurance target can be the performance.
For example, the effectiveness of the advertisement can be calibrated as the value of the duplicated data. 
If an advertisement company is assure that its opponents have duplicated its customer preference information, then the decreased amount of customers can be considered as the effect of the data breach.
Hence, the loss due to the data breach can be determined.

Insuring data or system performance can be adaptable for insurance contracts against data breaches.
Since the insured targets are immaterial, it is more convenient to cheat on it.
Indeed, when the users are not honest or when the users have incentives to misreport the missing data or system performance degradation, the risk sharing between the users and the insurers may be destroyed.
Nevertheless, dishonesty can be considered as a consequence of informational asymmetry between the users and the insurer.
We will elaborate on how to mitigate the asymmetry with the help of insurance contract modeling in Section \ref{sec:insurer observations}.

\subsection{Physical Layer Attacks}
We have mentioned previously that some APTs can target the critical infrastructures and the physical components in the OT system.
Here, we use the ransomware as a representative of physical layer attacks.

Ransomware originally refers to a class of malware that exploits security mechanisms to hijack user files and documents in computer systems for asking ransoms \citep{al2018ransomware}.
With the advances of the hackers, ransomware can also target the user ends in networked computer systems or even robots and unmanned-vehicles in autonomous systems. 
Conceptually, the attack target of a ransomware is making the systems unavailable to its users.
Therefore, we use it as an example of physical layer attacks.

Dates back to at least the floppy disk Trojan in 1989, ransomware has been advancing to extend the set of targets to infect and update the encrypting methods to secure the hackers profits.
Some variants include CryptoLocker \citep{liao2016behind} and Petya \citep{aidan2017comprehensive}.
It is predicted that the ransomware damage costs will exceed $265$ billion dollars by the year of 2031 \citep{RansomwarePrediction}.

One feature of ransomware is that it spreads over communication networks or communication links in Internet-of-Things (IoT) systems.
The monetary exchanges the attackers ask for is dependent on the number of disabled units in the system.
In this scenario, the cyber loss can be normalized to be the number of disabled computers. 
Accordingly, an intuitive and approach is to design the insurance to cover the ransoms for a fraction of disabled units.

The insured targets for cyber insurances against ransomware depend on the strategies for protecting the networked system. 
If the individual units in the network are federated, that is, they have independent firewalls and anti-malware software, then the insured targets are the individual computers.
In this scenario, individual users become the insurees. They have the choice to determine whether or not to participate in a cyber insurance contract.
If units in the networked system share the same protection technique, then the insured target should be the whole networked system.
In this scenario, the owner of the whole networked system is the insuree. 
The owner's decision on purchasing the cyber insurance ensures the capability of financially recovering from availability issues at the users' ends.

\section{Defense Mechanisms and Residual Risks}
\label{sec:defense mechanisms}
Aiming at defending the security of the systems, the users adopt various defense mechanisms keyed to individual attack vectors from the attackers.
In this section, we fist introduce generic models for describing the relationships between the defenders and the attackers to understand how to react to different types of attacks from the defenders perspective.
Then, we review different types of security investments that can aid the defenders to protect against the attacks mentioned in Section \ref{sec:attack models}.
Elaborating these investments reveals the potential uncovered cyber threats of certain systems used by various users, especially those from small to mid scale businesses and companies, and aids the design of the coverage plan of the cyber insurance.
Finally, we turn to the resiliency aspect of the systems by discussing the residual risks from cyber threats after the defense mechanisms have been set up.

\subsection{Modeling of Defense Mechanisms}
\label{sec:defense mechanisms:modeling}
To efficiently and successfully defend their systems, the defenders should be prepared for all kinds of attacks from all possible attackers.
This adversarial environment brings the necessity of figuring out potential malicious ones from a population of users.

Game theory is one of the emerging modeling techniques that digs into the rationality and strategic behaviors of individuals \citep{bacsar1998dynamic}.
The Nash equilibrium (NE) of a game characterizes the status of players' strategies where no one can benefit by deviating from the equilibrium. 
The strategic relations can be utilized to analyze the actions taken by rational agents when they face adversaries.
In particular, a simple two-player zero-sum game can explain the mental combat between an attacker and a defender.
For example, the work \citep{zhu2010heterogeneous} uses zero-sum games to model the relation between a system defender and an intruder in a networked intrusion detection system (IDS).
The intruder aims to scan the host machine for vulnerabilities while the defender monitors the suspicious behaviors taken place across the network and tries to detect any intruder.
The relation between a malicious attacker and an IDS has also been investigated under the two-person zero-sum game framework in earlier work such as \citep{zhu2009dynamic,alpcan2006intrusion,sallhammar2006stochastic}.

One of the most recognized application of game-theoretic tools in the domain of security is the Stackelberg security game \citep{sinha2018stackelberg}.
As a game with a special leader-follower structure, the Stackelberg game \citep{stackelberg1952theory} models the system defender as the leader who takes the defending action first and models the attacker as the follower who best-responds to the action of the defender.
Compared to a common simultaneous game where the players take actions at the same time without observing others' actions, a Stackelberg game adapts well to the scenarios of cyber threats and system attacks.
The reason lies in that all the investments of a defender to increase the security levels of the IT and OT are performed before the attacks.
Hence, the attackers can leverage the knowledge of the first-mover's action to determine their plays in the game.
Stackelberg games elaborate this timing.
Therefore, Stackelberg security games have been applied to many scenarios such as moving target defense \citep{fang2013protecting,feng2017stackelberg,pawlick2019game}, infrastructure security \citep{pita2008deployed,jain2010software,an2012protect}, proactive defense \citep{zhu2010stochastic,zhu2012hybrid,zhu2009game}, etc.

Information plays an important role in the strategic relations between the defender and the attacker.
When the defender has to figure out the true target of the attacker or to find the stealthy malicious user from a population of users before adopting any defensive action, Bayesian games offer the tools to deal with the incompleteness of information in the environment.
Instead of the scenario described by the common knowledge assumption in complete information games, what we are dealing with in real life is a slightly more complicated scenario where the information of the game or of other players are incomplete.
Harsanyi's Bayesian game \citep{harsanyi1967games} introduces the notion of types, which describes the state of a player's mind when one constructs beliefs of the state of the game using observable information.
Due to the modeling power, Bayesian games have been applied to defensive cyber deception \citep{huang2019dynamic,pawlick2021game,zhu2019game}, defenses against APTs \citep{huang2019adaptive,huang2020dynamic,zhu2018multi}, CPS security \citep{pawlick2015flip,pawlick2017strategic,huang2020dynamic}, etc.

Apart from game theory, another meaningful model suitable for applications involving informational incompleteness is the partially observable Markov decision process (POMDP).
While Markov processes naturally applies to dynamic environments such as encrypting probabilities for attackers to appear into the transition kernel , the decision maker's partial observation can model the the defender's knowledge about whether a user is classified as benign or malicious.
For example, the authors in \citep{miehling2018pomdp} adopt POMDP to study the dynamic defense of cyber networks where the defender has partial observation of the security state of the system and has to construct belief of the security state using information from security alerts.
In \citep{kurt2018online}, the authors use POMDP models to investigate the dynamic detection of cyber threats where the defender has partial information about the security of the environment.
Related modeling techniques can also be found in \citep{zhu2020cross,sarraute2012pomdps,mc2016data}.

\subsection{Types of Security Investments}
\label{sec:defense mechanisms:investments}
In the popular game models for characterizing the strategic relations between the system defender and the attackers mentioned in Section \ref{sec:defense mechanisms:modeling}, the defenders' defending strategies vary according to the attack models.
From the user's perspective, this variety captures the difficulty in finding a unified approach to strengthen system security.
In the following, we elaborate on the possible security investments for the attack models discussed in Section \ref{sec:attack models}.
Due to the differences in the expenses and the difficulties of deployments of the security investments, network users are recommended to invest in a reasonable amount that is feasible for them to protect against corresponding attacks.
If certain security investments are too expensive for the users, they can resort to cyber insurance to make their systems resilience.

\paragraph{Against Insider Threats}
At a first glance, employee training may be a effective investment to defend against insider threats.
However, it can only benefit businesses and companies when the insiders are unintentional perpetrators.
The reason lies in that the other two types of insiders, namely the masqueraders and the traitors, possess malicious intentions.
Hence, they can be well-behaved according to certain employee behavior standards but they can still stealthily harm the system. 

The intricacy of insider threats lie in the fact that we can hardly trust anyone in the system \citep{colwill2009human}.
In fact, a number of game-theoretic frameworks to address the insider threats have utilized a probabilistic way to model insiders.
That is, any player in a game can become an insider and create insider threats to the system with some probability.
Examples of these frameworks include \citep{joshi2020insider,huang2021duplicity,zhu2012guidex,casey2016compliance}.

This consequence of insider threats brings the notion of zero trust \citep{garbis2021zero}.
The zero trust architecture is a cybersecurity paradigm which emphasizes that trust must be evaluated continuously \cite{rose2020zero}.
This infers, for example, that even when an individual has been granted access to certain enterprise data or resources, the system will not trust this person anymore as soon as the one-time access expires.
The zero trust architecture contains more basic tenets that benefits the cyber systems in the protection of privacy, the allocation of resources and computation power, the authorization of individuals or entities, etc \citep{stafford2020zero}.

Adopting the zero trust security paradigm will increase the overall security level of cyber systems. 
However, as a more sophisticated and comprehensive security structure, the zero trust architecture certainly costs much more to implement.
A simple reason is that to enforce zero trust, every action in the ICT needs to monitored.
The dynamic nature requires more expenses to be invested.
Therefore, despite the advances of zero trust security, efforts are still needed to enable its implementability in practice and compatibility with the investment power of small scale businesses.

\paragraph{Against APTs}
Defending against APTs is a challenging task since APTs combine different attack vectors and can last for a reasonably long period.
The APT defense method can be divided into mainly three types \citep{alshamrani2019survey} due to the many-phased attack process APTs possess.
The three types include monitoring, detection, and mitigation.

The monitoring can be classified as a screening or filtering of relative data prior to the detection of threats.
Business owners may need to build large-scope monitoring systems that cover the entire ICT and monitor all the areas from the memory to the codes.
This comprehensive approach will certainly generate an enormous amount of expenses.
However, since APTs can perform their initial invades at arbitrary entry points of a system, this type of monitoring is necessary.

While the monitoring utilizes traditional techniques, the detection of APTs benefits more from recent technical advances.
For example, machine learning based APT detection systems \citep{siddiqui2016detecting,ghafir2018detection,chandran2015efficient}, reinforcement learning based APT detection \citep{huang2019adaptive,huang2019deceptive}, and game-theoretic detection frameworks \citep{huang2020dynamic,yhuang2020dynamic,rass2016gadapt,zhu2018multi}.

In recent years, proactive methods in mitigating damages caused by APTs emerge.
The engagement of defensive deception techniques and moving target defense systems, among the other proactive methods, have proved their effectivenesses \citep{zhu2018game}.
Implementation of honeypots, as an example of defensive deception, can have the chance to trap the hackers into a carefully designed trap, so that the defender is able to learn the invading strategies and attack targets of the hacker \citep{huang2019adaptivehoneypot,pawlick2019game,huang2021duplicity,huang2022reinforcement}.
Moving target defenses, as a dynamically changing mechanism, protect the system by creating an ever-changing view for the attacker.
Dynamic game is one of the promising tools for designing moving target defense systems \citep{zhu2013game,pawlick2019game}.

Investments on novel technologies can improve the successful detection rates and enhancing cyber security by mitigating potential harms.
However, technologies update very rapidly.
Knowing how to choose the correct and compatible method can sometimes become a challenging task for businesses.

\paragraph{Against Data Breaches}
Although there exist efforts in the research of data breach incident predictions \citep{sarabi2016risky}, data breaches are hard to avoid because of the various initiatives that can lead to the leak of information.
One factor that serves as the start point of many data breach incidents is phishing. 
Hence, employee training is the very first investment businesses and companies can think of to defend against data breach threats.
The training can simply contain contents such as methods to recognize phishing emails or messages so that even some of the emails pass through the automatic classification system, they can be considered as malicious by human eyes.

The second straightforward investment is maintaining backups of data and information or separating the storage into partitions and save them in a disjoint fashion.
Since the backups of data are often saved in certain devices that are offline or disconnected with the main communication networks, they can be unreachable from hackers.
Though keeping backups cannot stop the data stored or used in the ICT to be stolen, they can at least secure the normal functionalities of the systems when data breaches happen.
Partition of storage is also away to reduce the amount of data being hacked.
Hence, it reduces the losses due to data breaches to the least.

In recent years, multi-factor authentication methods \citep{ometov2018multi} emerge to replace traditional authentication methods which contain only a single factor.
It is realized that single-factor authentication is not adequate against many threats \cite{gunson2011user}.
Instead, by coupling one factor of identification data such as username and passwords with another factor of personal belongings such as smart phones, the two-factor authentication methods have already enriches the authentication procedure and have enhanced the accuracy and security of authentication. 
In multi-factor authentication methods, biometric factors enabled by modern technologies such as natural language processing and facial recognition systems have become a reliable data source for authentication purposes.
These biometric factors include, but not limited to, personal voice characteristics, facial information, fingerprints, and hand geometry \citep{ometov2018multi}.

Encryption of data provides an additional layer of protection which adds to the previous security investments.
By encrypting the sensitive information, even when the previous security mechanisms fail, the hackers may be unable to read the encrypted data and they have to stop before harming the system further.

\paragraph{Against Ransomware Attacks}
Ransomware, as a special malware which can spread through the ICT and other networks connecting autonomous systems, can be successfully blocked by generic anti-virus software and firewalls \cite{aslan2020comprehensive}.
As a more fatal malware that can completely shut down the services of businesses or stop the functionality of companies, more detection systems and defense frameworks that are explicitly designed for ransomewares appear in recent years \citep{el2018intrusion,wan2018feature,almashhadani2019multi,zhao2021combating}.

\subsection{Residual Risk and Its Connection with Cyber Insurance}
\label{sec:defense mechanisms:residual}
Despite the effectiveness of the defense mechanisms introduced in Section \ref{sec:defense mechanisms:investments}, hoping to gain complete invincibility against cyber threats through security investments is impossible.
Indeed, attackers can always seek the vulnerabilities from current defending mechanisms and try designing specific attacks targeting these weaknesses.
Therefore, cyber risks always exist in the long run.
That being the case, addressing the challenges in cyber risk assessment will be crucial for building a resilient ecosystem. 

Due to the sophistication in the interdependency among the attack vectors in different layers and the uncertainties present in the quantification of the reduction of cyber risks caused by defense methods, a probabilistic description of the relation between the security investment and cyber risks is plausible.
Namely, security investment can reduce cyber risks but can never eliminate them.
The remaining risk passing through the functions of defending mechanisms is referred to as the residual risk.
The residual risk is one of the central concerns of cyber resilience, for enhancing cyber resilience can be equivalently understood as mitigating the residual risk.

In a practically meaningful stochastic relation between the security investment and the residual risk, a parametric probability distribution is one of the most concise and convenient description.
Intuitively, increased security investment should be able to at least reduce the probability that cyber attacks successfully harm the system.
This relation can be precisely described by the notion of stochastic dominance \citep{levy1992stochastic}.

\begin{figure}
\centering
%\subfigure{\includegraphics[width=0.44\textwidth]{dominance1.eps}
   % \rule{4cm}{3cm}
    %\label{fig:subfig1}
%}
%\subfigure{
\includegraphics[width=0.74\textwidth]{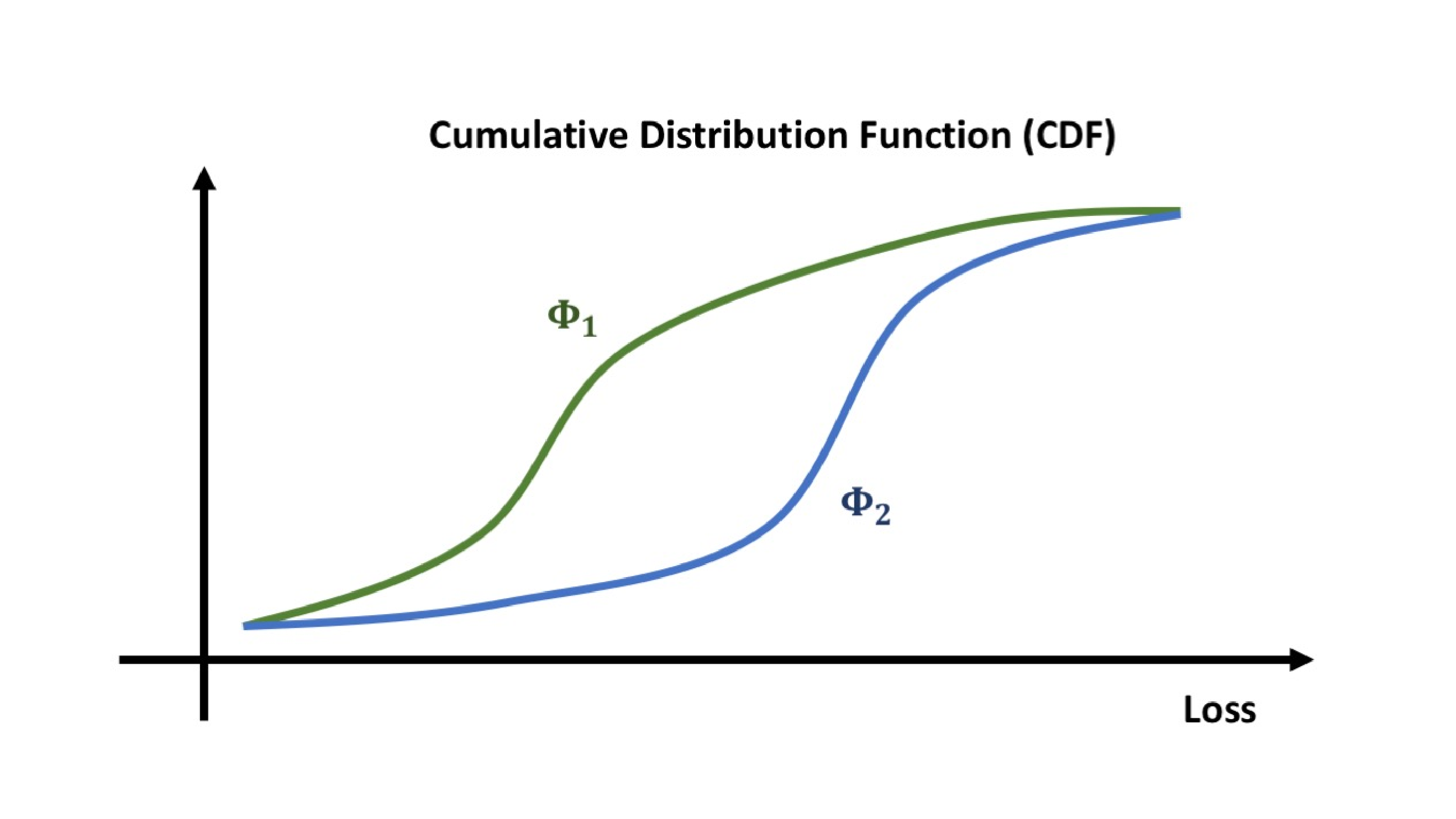}
   % \rule{4cm}{3cm}
    \label{fig:subfig2}
%}
\caption{First-order stochastic dominance relation viewed from the CDF.
\label{fig:dominance}}
\end{figure}

Consider two random variables $Z_1$ and $Z_2$.
Let $\phi_1$ and $\phi_2$ denote the probability density functions (PDFs) of $Z_1$ and $Z_2$, respectively.
Let $\Phi_1$ and $\Phi_2$ denote the cumulative distribution functions (CDFs) of $Z_1$ and $Z_2$, respectively.
Then, $Z_1$ stochastically dominates $Z_2$ in the first order if $\Phi_1 \leq \Phi_2$.
Consider $Z_1$ and $Z_2$ to be the residual risks of a cyber system under different amount of security investments.
Then the stochastic dominance relation can be intuitively understood as that, compared to $Z_2$, $Z_1$ contains greater likelihoods for severer cyber losses to happen.
In this scenario, the investment that leads to $Z_2$ should be higher than the investment that leads to $Z_1$.
An example of the relations %between the PDFs $\phi_1$ and $\phi_2$ and
between the CDFs $\Phi_1$ and $\Phi_2$ is illustrated in Fig. \ref{fig:dominance}.

Parameterized probability distributions and the stochastic dominance relations enables the insurer to calibrate the residual risk when the users' security investments on protecting their cyber systems are observable to them.
Using these mathematical tools, the effects of defense mechanisms described in Section \ref{sec:defense mechanisms:investments} can be holistically described in a formal way.
Combining the modeling of the users' monetary utilities or costs, these relations enables the insurer to evaluate the profits from claiming a cyber insurance contract with certain premium and coverage.

In the sequel, we will elaborate on the available observations of the insurer, based on which she aims to design cyber insurance contracts.
Then, we will introduce the cyber insurance design framework.

\section{Insurer's Observations and the Principal-agent Model}
\label{sec:insurer observations}

\subsection{User Behavior Monitoring}
Users obtain financial supports and gain protection against cyber losses from cyber insurances because of the risk sharing.
However, the loss coverage provided by the insurance contract is not effective until the user makes a prior payment to the insurer as the premium.
That is, an insurance company is essentially a business organization.
This means that the existence of the cyber insurance market depends on whether insurance contracts produce profits.

Since a cyber insurance contract is offered to the users by the insurer, the insurer can carefully design the insurance plan to maximize her profits.
One essential reference for designing the insurance contract is the insurer's observation of the users' security investments.
The calibration of the relation between the coverage and the premium is often performed using historical data of the users' willingness to invest on system protection and the successes of cyber attacks targeting the systems.
While these data cannot predict what the users will do after they purchase the insurance plan, they can, to some extent, help the insurance company learn the effectiveness of various security investments against different cyber threats.
Utilizing models such as the parameterized probability distributions introduced in Section \ref{sec:defense mechanisms:residual}, the insurer can obtain a clear picture of the potential cyber risks.
Hence, they can design the proportion of losses to cover and the expense to charge the users.

The behavior of the users can change after they are insured.
This makes the monitoring of the users' behaviors very important.
For example, consider an owner of a networked system.
Suppose that there is an offer of a cyber insurance contract that fully covers the losses due to cyber threats with a reasonable amount of monetary charge.
Then, if the owner purchases this insurance plan, she will not have any incentive to invest on system protection.
Since whether or not she updates the firewalls or deploys IDSs, the losses will be fully covered by the insurance company. 

The above phenomenon is referred to as the moral hazard issue in the economic literature \citep{pauly1968economics}.
More formally, moral hazard describes a situation where individuals tend to behave more recklessly when one does not fully bear the risk. 
It appears when there is information asymmetry between the users and the insurer.
Insurance contacts are severely affected by moral hazard.
The reason lies in that if the users does not follow the statistical pattern of historical behaviors and become reckless after they are insured, the design of the contract can fail to function and can lead to negative profits to the insurer.
In the following, we will formally describe the standard framework for insurance design and how a variation of the framework addresses the moral hazard issue.

\subsection{Principal-Agent Problems}
\label{sec:insurer observations:p a problem}
A suitable framework for cyber insurance design is a class of principal-agent (P-A) problems \citep{grossman1992analysis}.
The P-A problem can be viewed as a bilevel optimization problem \cite{dempe2002foundations}, where the principal solves the upper-level optimization problem and the agent solves the lower-level problem.
P-A problem has been applied to the modeling of the relations between, for example, employers and employees, buyers and sellers, citizens and elected officials, etc.
Prior works using the P-A framework to design insurance contracts include \citep{marotta2017cyber,khalili2018designing,bohme2010modeling,zhang2017bi,zhang2019mathtt}.

We proceed to formally introduce two types of P-A problems. The first one correspond to the scenario where the insurer can perfectly observe the action of the user after they are insured. The second one deals with the opposite situation where the moral hazard issue occurs and a calibration of the user's action is required.

\begin{figure}
\centering
\includegraphics[width=0.8\textwidth]{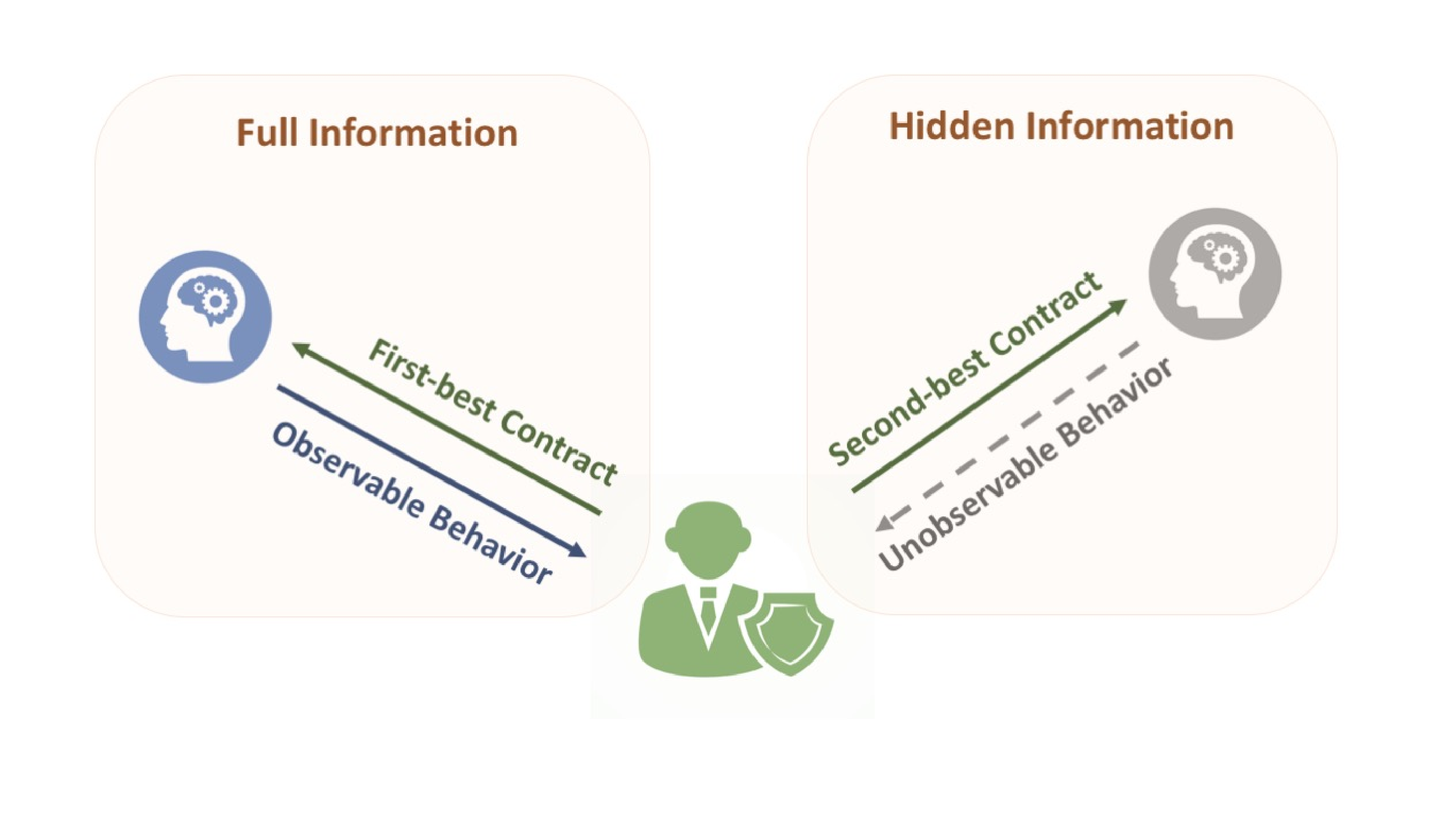}
\caption{Full-information and hidden-information P-A problems for cyber insurance design.
\label{fig:triad}}
\end{figure}

Let $x\in X$ denote the security investment from the user.
In practice, $x$ can refer to any type of investments (as described in Section \ref{sec:defense mechanisms:investments}) depending on the potential cyber attacks that can occur.
Let $\xi$ denote the residual risk under certain deployed defense mechanism. 
The residual risk is a random outcome from a probability space $(\Xi, \mathcal{F})$.
The stochasticity of the residual risk is described by the parameterized probability distribution $P(\xi,x)$ which depends on the security investment.
The insurer's profit by designing the insurance contract $w: \Xi \rightarrow \mathbb{R}$ is described by the utility function $V:\Xi\times \mathbb{R}\rightarrow \mathbb{R}$, which also depends on the residual risk $\xi$.
The user's cost from the security investments and loss due to the residual cyber risks after enrolling in the insurance contract is described by the disutility function $U:X\times \mathbb{R}\rightarrow \mathbb{R}$.

When the insurer has perfect observation of the user's actions of security investments, the full-information P-A problem can be formulated as:
\begin{equation}
    \begin{aligned}
         \max_{w(\cdot), x}
         & \int_{\Xi}V(\xi,w(\xi)) dP(\xi,x) \\
         \text{s.t. } \ \ 
         &\int_{\Xi}U(w(\xi),x)dP(\xi,x)\leq \Bar{U}, \text{(IR)},
         \label{eq:full info contract problem}
    \end{aligned}
\end{equation}
where $\Bar{U}$ denotes the minimum cost of the user when she is not enrolled in the cyber insurance contract.
In the optimization problem (\ref{eq:full info contract problem}), the IR constraint refers to individual rationality \citep{holmstrom1979moral}  which guarantees beneficial participation.
The reason why the insurer in problem (\ref{eq:full info contract problem}) possesses full-information of the user's action lies in that the security investment is fully controlled by the insurer.
That is, as long as an investment satisfies IR, it is feasible for the optimization problem (\ref{eq:full info contract problem}).
The insurer can choose a security investment for the user to obtain optimal profit.
This can also be observed from the single-level structure of problem (\ref{eq:full info contract problem}).

When the insurer cannot observe the action of the user, the moral hazard issue may occur.
In this scenario, to design a profitable insurance contract, it is necessary for the insurer to calibrate the user's action.
The hidden-information P-A problem, also referred to as the hidden-information moral hazard problem, can be formulated as:
\begin{equation}
    \begin{aligned}
         \max_{w(\cdot), x}
         & \int_{\Xi}V(\xi,w(\xi)) dP(\xi,x) \\
         \text{s.t. } \ \ 
         &\int_{\Xi}U(w(\xi),x)dP(\xi,x)\leq \Bar{U}, \text{(IR)}, \\
         & x\in \argmin_{x'\in X}  \int_{\Xi}U(w(\xi),x')dP(\xi,x'), \text{(IC)}. 
         \label{eq:hidden info contract problem}
    \end{aligned}
\end{equation}
The difference between problem (\ref{eq:hidden info contract problem}) and problem (\ref{eq:full info contract problem}) lies in the additional IC constraint in problem (\ref{eq:hidden info contract problem}), which refers to incentive compatibility \citep{holmstrom1979moral}.
By adding the IC constraint, problem (\ref{eq:hidden info contract problem}) takes into account the user's rationality.
In other words, since the insurer cannot observe the action of the user and has no control over her security investment, the insurance plan that the insurer designs has to stand in the user's place and minimize the monetary cost of the user.
Furthermore, the IC constraint adds an additional layer of decision making and makes problem (\ref{eq:hidden info contract problem}) a bilevel optimization problem.

From an optimization-theoretic perspective, the optimal insurance contract obtained by solving problem (\ref{eq:full info contract problem}) is referred to as the fist-best contract, and the optimal insurance contract obtained by solving problem (\ref{eq:hidden info contract problem}) is referred to as the second-best contract.
The insurer's profit under the first-best contract cannot be worse than the profit produced by the second-best contract.
The reason lies in that the feasible set of (\ref{eq:hidden info contract problem}) is a subset of the feasible set of (\ref{eq:full info contract problem}).

From a practical perspective, the first-best contract suffers from the moral hazard issue.
Hence, (\ref{eq:hidden info contract problem}) is a more reasonable choice for the design of cyber insurance contract in practice.
The practicality of (\ref{eq:hidden info contract problem}) is obtained at the cost of a deduction in the insurer's profit.

As a bilevel optimization problem, (\ref{eq:hidden info contract problem}) is challenging to solve both analytically and numerically (\cite{dempe2002foundations}).
One approach to characterize the optimal insurance contract of problem (\ref{eq:hidden info contract problem}) is to use the first-order optimality condition of the IC constraint and transfor the bilevel optimization problem to a single-level optimization problem.
Critical factors that enables the first-order approach include the shapes of the utility functions $U$ and $V$, or, the risk sensitivities of the insurer and the user, which we are going to elaborate in Section \ref{sec:risk modeling}.

In practice, the insurance contract $w(\cdot)$ is often restricted to the class of linear insurance plans consist of a premium payment $\mathfrak{p}>0$ and a coverage rate $\mathfrak{c}\in(0,1)$.
The linear insurance plan is commonly used since it is convenient and intuitive.

\section{Modeling of Risk Preferences}
\label{sec:risk modeling}
\begin{figure}
\centering
\includegraphics[width=0.8\textwidth]{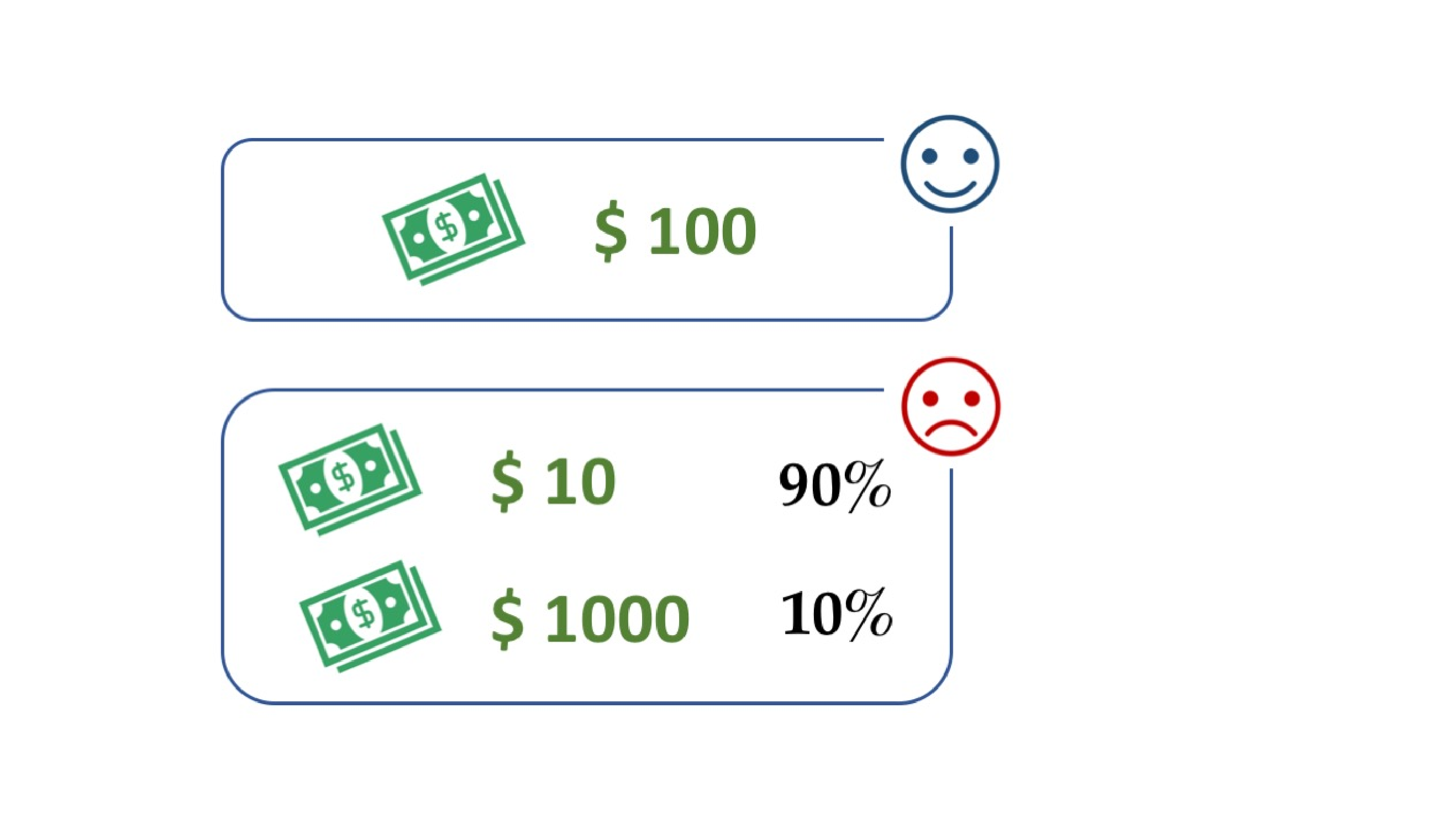}
\caption{A risk-averse decision-maker prefers the  certain event on the top to the random event on the bottom whose expected value is higher.
\label{fig:riskaverse}}
\end{figure}

Risk preference refers to the decision-maker's perception of losses or gains in an environment with uncertainties.
This perception influences whether one quantifies probabilistic events as valuable or not.
There are three common types of risk preferences, namely, risk-neutral, risk-averse,and risk seeking.
Risk-neutral decision-makers evaluate probabilistic events purely according to its law.
That is, they consider a certain event and an uncertain event as equivalent when they are equal in the average sense.
A risk-averse decision-maker  tends to dislike or avoid events that are uncertain.
Sometimes, they prefer a certain event to an uncertain event even when the former benefits them less than the latter in the expected sense (see Fig. \ref{fig:riskaverse}).
A risk-seeking decision-maker holds the opposite preference compared to a risk-averse decision-maker.

One of the most consequential impact risk preferences enforces on cyber insurance is the existence of the cyber insurance markets.
It has been discovered that the cyber insurance market only exists when the insurer is risk-neutral and the users are risk-averse \citep{khalili2018designing,bohme2010modeling,marotta2017cyber}.

In the sequel, we will briefly introduce the modeling of risk preferences. Then, we will explain how risk preferences can help enhancing cyber resilience.

\subsection{Risk Modeling}
\label{sec:risk modeling:modeling}
The modeling of risk preference has a long history.
It dates back at least to the expected utility theory of \citep{von2007theory}.
According the hypothesis of expected utility theory, rational agents make choices based on the expected utility they perceive.
The risk appetite of agents are captured by the utility functions associates with the agents.
When the utility functions are concave, a profit-maximizing agent exhibits risk aversion.
The famous Arrow-Pratt measure of absolute risk aversion \citep{arrow1971theory,pratt1978risk} extends the definition of risk aversion merely based on the second derivative by involving the first order information of the utility function to make the risk preferences invariant with respect to affine transformations.
The choice of utility functions heavily affects the properties of the contracts obtained by solving the P-A model.
For example, risk-aversion is essential in proving the monotonicity of optimal contracts \citep{stole2001lectures}.

Risk preferences appear because of the uncertainties in the environment.
The expected utility expression mainly evaluates how the randomness in the uncertainties affect the decision-maker's choice when the subjective probability rule is fixed.
The cumulative prospect theory in \citep{tversky1992advances} introduces an additional layer in the quantification of randomness, which is the perception of probabilities.
In particular, \citep{tversky1992advances} uses experimental and psychological evidences to show that the perception of probability is also objective.
Hence, in addition to agent-specific utility functions, \citep{tversky1992advances} considers probability weightings when the expected utilities of decision-makers are specified.
This comprehensive expression of preference enriches the original approach where only the utility function varies according to individuals.
The idea of probability weightings can be influential to cyber insurance design.
For example, the stochastic relation between the security investments from the users and the law of the cyber loss can be equivalently understood as a probability distortion.
Then, with the probability weighting associated with individual risk perceptions, the physical impact of the investments to the cyber losses and the psychological impact of individual preference to the cyber losses may be unified and expressed holistically.

A modern approach to modeling individual risk preferences is the use of risk measures.
In the seminal work \citep{artzner1999coherent}, the authors have introduced the notion of coherent risk measures (CRMs) to quantify uncertainty.
The coherency refers to the four axioms a risk measure has to satisfy, namely, monotonicity, sub-additivity, translation invariance, and positive homogeneity.
These axioms not only rigorously define how to order or compare different random quantities, but also take into account the practicality that CRMs needs to adapt to in scenarios such as investment and asset management. 
For example, the sub-additivity axiom is closely related to the hedging effect in finance, which states that an investment position can be used to offset the potential losses by an adverse one.
Since the introduction of CRMs, there has been a trend on studying both the theoretical properties and the applications of CRMs \citep{follmer2016stochastic,ruszczynski2006optimization,pflug2014multistage,bertsimas2015optimizing,liu2020robust,noorani2022embracing}.

Among the popular CRMs, average value-at-risk (AV@R) \citep{rockafellar2000optimization}, which also bears the name of the conditional value-at-risk, the expected shortfall risk, and the expected tail loss, is one of the most investigated.
The reasons lie in but not limited to the following.
Firstly, it is a direct extension of the popular non-coherent value-at-risk (V@R) measure.
V@R has been used as the industrial standard before the introduction of coherency.
Secondly, the Kusuoka representation \citep{kusuoka2001law} indicates that any law-invariant CRM can be represented by a class of AV@Rs.
Thirdly, the computation of AV@R can be equivalenty cast into a convex optimization problem \citep{rockafellar2000optimization}.

Adopting CRMs also adds robustness to the agents' decisions.
This can be seen from the dual representation of a CRM \citep{shapiro2021lectures}.
In particular, the dual representation of a CRM coincides with a distributionally robust stochastic programming problem \citep{rahimian2019distributionally}.
The distributional robustness can be inferred from the fact that, by adopting a CRM, the decision-maker is actually making decisions under the worst-case scenario probability distribution.

The modern CRM approach to risk modeling has its advances in both the convenience of mathematical expressions and the richness in describing objective preferences.
We will introduce how CRMs can be applied to the design of cyber insurance contracts and how CRMs contribute to cyber resilience in the next section.

\subsection{Enhancing Cyber Resilience}
\label{sec:risk modeling:enhance}
As we have mentioned in Section \ref{sec:insurer observations} that the cyber insurance market only exists when the insurer is risk-neutral and the user is risk-averse.
In fact, the existence of the market does not necessarily mean that it is a beneficial market.
In the literature, researchers have discovered the phenomenon that the cyber insurance market will decrease the security level of networked cyber systems \citep{khalili2018designing,bohme2010modeling,marotta2017cyber}.
The reason lies in that the market disincentivizes the users to invest on system protections.
This phenomenon leads to a worse-off social welfare, though individual users receive reduced overall costs with the help of cyber insurance.
Note that the cause of this phenomenon is not the moral hazard issue, since the insurance models considered in \citep{khalili2018designing,bohme2010modeling,marotta2017cyber} do contain the incentive compatibility constraints.
This drawback of the cyber insurance market have prevented its practical implementation.

\begin{figure}
\centering
\includegraphics[width=0.8\textwidth]{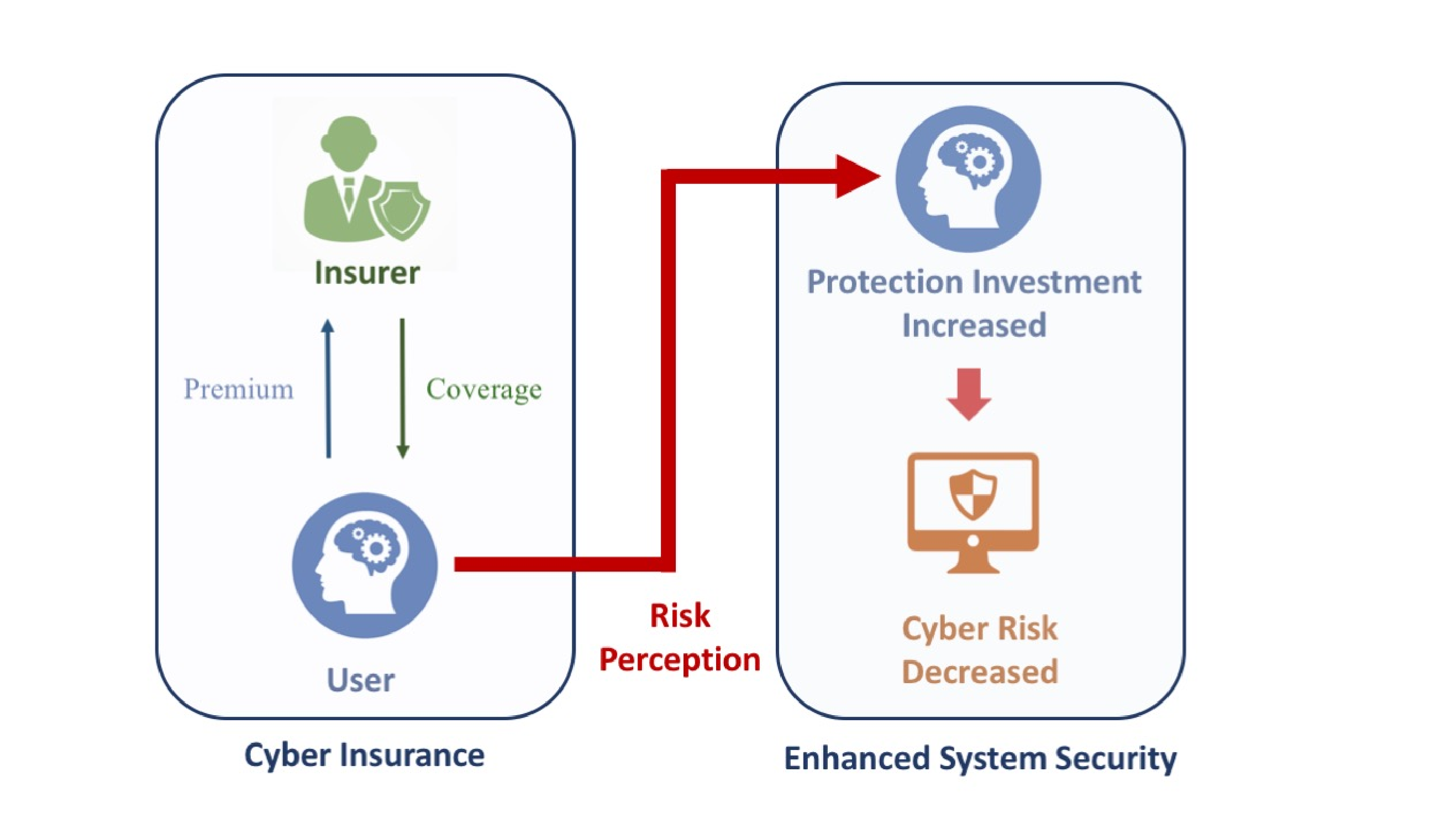}
\caption{Risk perception is the key to make cyber insurance beneficial not only for cyber resilience but also for cybersecurity.
\label{fig:enhance}}
\end{figure}

Among the other approaches, the recent work \citep{liu2022role} circumvents this challenge by enriching the risk perceptions considered in the classic cyber insurance frameworks that build on the P-A model (See Fig. \ref{fig:enhance}).
The authors of \citep{liu2022role} have shown analytically that, with properly specified risk perceptions, purchasing cyber insurance can enhance the security level of cyber systems by incentivizing the users to increase their security investments.
In particular, \citep{liu2022role} requires that the insurer's risk perception should be more sensitive to the stochastic dominance shift induced by the user's action than the user's risk perception does.
This investigation into the evolution of the risk attitudes according to the variation in the user's investment action is enabled by adopting the risk measures.

While the work \citep{liu2022role} has shown that cyber insurance, as a means to increase cyber resilience, can also enhance the level of cybersecurity, it only considers linear cyber insurance plans.
In the next section, we will investigate the position of risk preferences in cyber insurance with a generic insurance contract and consider cybersecurity and cyber resilience holistically.

\section{Insurance Design with Preference Manipulation}
\label{sec:preference manipulation}
We have discussed previously that defense mechanisms are deployed for potential cyber attacks to increase the level of cybersecurity while cyber insurance is the means to enhance the resilience of systems in order to guarantee post-damage recovery.
On the one hand, cyber insurance is an independent consideration for businesses whose security investment is fixed or takes a long time to upgrade.
On the other hand, companies who update their system protection methods constantly will more likely consider the investment in insurance and defense mechanism as a whole.
In this scenario, a prominent phenomenon that may occur and affect the investment decisions is the moral hazard issue.

When the moral hazard issue happens, users tend to invest less on system protection after their cyber losses are covered by cyber insurances.
While this pattern is out of the rationality of the users, it can significantly decrease the security investments and leads to poor system protection.
Furthermore, in scenarios of moral hazard, the first-best insurance contract is not practical anymore.

As we have discussed in Section \ref{sec:insurer observations:p a problem}, the hidden-information P-A problem is one approach to designing a cyber insurance contract when moral hazard cannot be ignored.
Although the second-best contract obtained from (\ref{eq:hidden info contract problem}) does not suffer from the moral hazard issue, it generates a lower profit to the insurer than that of the first-best contract of (\ref{eq:full info contract problem}).

Recently, in \citep{liu2022mitigating}, the authors have proposed using risk preference design to mitigate the moral hazard issue.
The notion of risk preference design is motivated by the fact that human risk preferences are not stable \citep{schildberg2018risk,anderson2009risk}.
In particular, the authors suggest using approaches such as  information distributing and nudging to shape the risk preferences of potential insurees.
This preference manipulation can help the insurer in obtaining a special state of a population where individual risk preferences benefits the insurer in her design of the insurance contract at a certain monetary cost.
The main benefit that the insurer can obtain through risk preference design is the mitigation of the intensity of the moral hazard issue, which is measured by the difference from the user's action solving the full-information P-A problem to the one the user actually chooses given the first-best contract.

One of the most significant consequences of the mitigation of the moral hazard issue is that the first-best insurance contract becomes more applicable in practice.
Since, by designing the risk preferences, the users may not deviate from the actions specified by the full-information P-A problem after they have been covered by the cyber insurance, the insurer has a better chance to obtain the first-best profit other than the second-best one.
This advantage creates motivations for the insurer to expand the cyber insurance markets to cover potential cyber losses of various classes and types.

Another notable contribution of \citep{liu2022mitigating} is the proposed holistic framework for the joint design of cyber insurance and risk preferences.
Built on the classic P-A problems, the framework of \citep{liu2022mitigating} provides a convenient way to analyze the effect of risk design and insurance design.
What the insurers need to do other than designing a traditional cyber insurance contract is calibrating the monetary cost of preference shaping.
This cost could include the expenses in distributing certain information, in advertising, and in hiring expertise in elicitating users' risk preferences.

\begin{figure}
\centering
\includegraphics[width=1\textwidth]{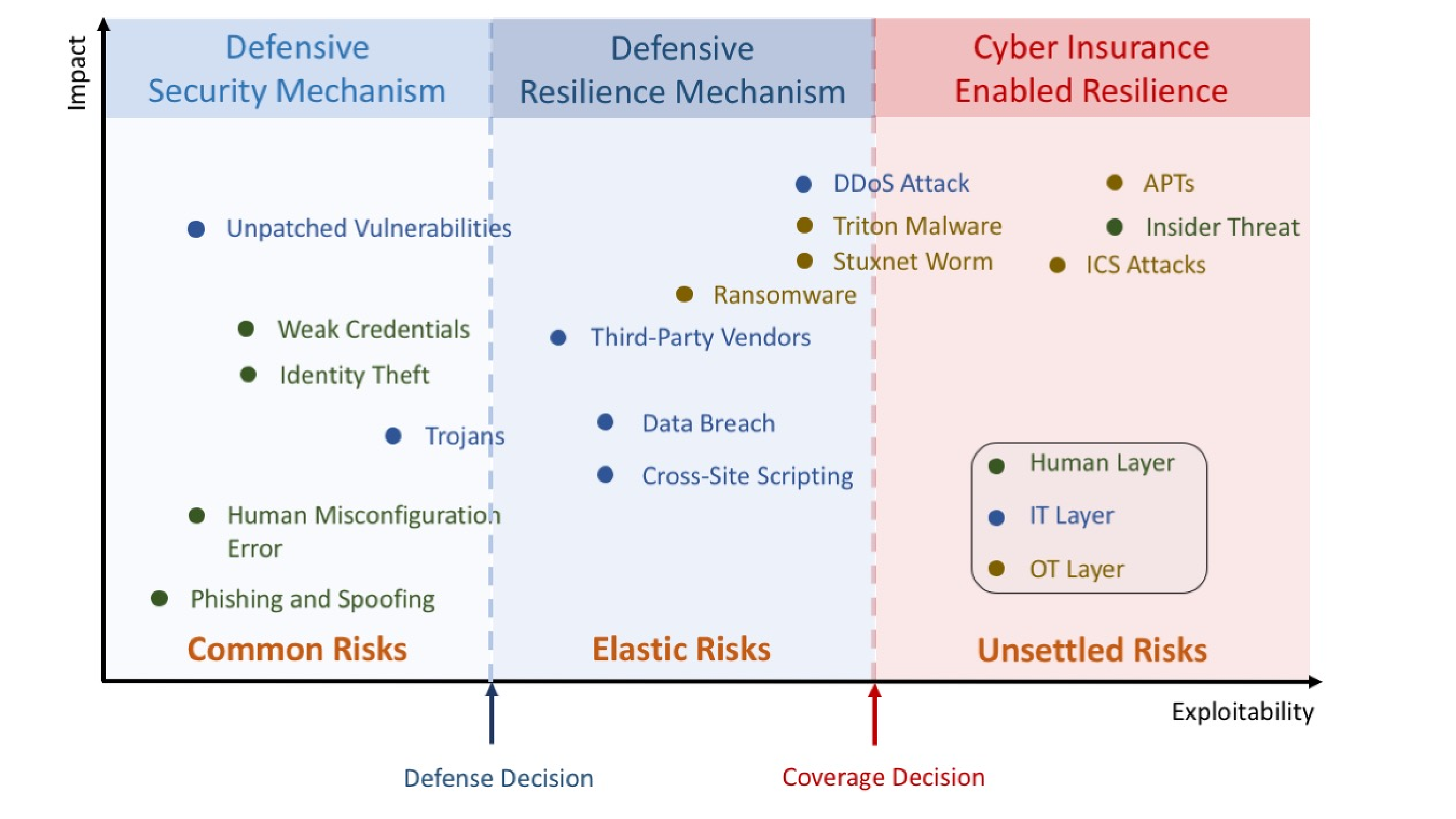}
\caption{Categorization of cyber threats. Cyber insurance leads to cyber resilience by enabling post-damage recovery from unsettled risks.
\label{fig:coverage}}
\end{figure}

Preference manipulation benefits cyber resilience in the following ways.
Firstly, the insurer can directly design the risk preferences of the users so that they become more careful in taking actions and more risk-averse in planning their defense strategies.
Then, these users are less likely to behave recklessly and hence cyber resilience is enhanced due to upgraded defense methods and increased security investments.
The cautiousness in the users' preferences can help maintaining a reasonable level of cyber resilience even under mis-assessed cyber risks, since it is likely that these users have set up necessary security devices and strategies in the ex-ante stage.
Secondly, the reduction of moral hazard enhances cyber resilience due to the following reason.
As we have mentioned in Section \ref{sec:insurer observations:p a problem}, the hidden-action P-A problem is challenging to solve due to its bilevel structure.
So, an approximation of the second-best contract is often a more feasible anticipated solution.
This fact indicates that an insurance contract obtained by taking into account the exact incentives of the users is hard to obtain.
However, due to the mitigation of moral hazard, the incentives of the user and the insurer become more aligned.
Hence, even though we only have access to an approximate second-best insurance contract, it still characterizes, to a reasonable extent, the incentive of the users.
Therefore, the users' actions will not deviate too much from the anticipated actions of the insurer, which leads to a resilient design of cyber insurance.
Thirdly, risk design enhances cyber resilience by pushing the border of the coverage.
Since an increased level of risk-aversion makes users invest more on system protection and security, common threats can be taken care of by the users themselves. 
This leaves more space for the insurer to focus on those attacks that requires more expertise to defend against, which results in a higher level of resiliency (See Fig. \ref{fig:coverage}).

\section{Dynamic Insurances}
\label{sec:dynamic}
In previous sections, we have focused on the scenario where the participation in a cyber insurance contract is modeled as a single stage decision problem.
In other words, the users only decide once whether or not to purchase the cyber insurance.
However, in practice, cyber threats always exist and they upgrade themselves constantly.
For example, one of the features of an APT is that the attacker will always adopt the state-of-the-art methods to achieve his targets.
This evolving environment motivates the consideration of dynamic cyber insurance contract, whose coverage plan is not fixed but changing according to both the security states of the networked environment and the types of attacks of cyber threats.

The frequency of updating the cyber insurance contracts should depend on the cyber landscape.
In particular, the future landscape of IT systems can be influenced by the advances of both the attack vectors and the defense techniques.
When the upgrades of the attack vectors lead the updates of the defense techniques, we would more likely set the a higher premium for the insurance. 
When the defense mechanisms are more advanced, we probably can deploy a high coverage since the residual risk is not severe.
In both of the cases, recalibrations of the probabilities and the severities of cyber losses are needed, for they provide necessary statistics for the execution of insurance contracts.

The extension of the P-A problem to its dynamic setting can aid the design of dynamic insurances \citep{crawford1985dynamic}.
One of the foundations of dynamic insurance design is dynamic cyber risk assessment \citep{badhwar2021dynamic}.
In \citep{zhang2021optimal}, the authors have adopted a Markov decision process to capture the dynamic correlations between the cyber risks and the users' decisions.
The Markov decision process is later incorporated into a P-A problem to aid the design of dynamic cybere insurance.
In \citep{papastergiou2018mitigate}, the authors have considered dynamic supply chain cyber risk assessment from an experimental perspective.
While they have focused on the mitigation of supply chain cyber risks, the methodology of \citep{papastergiou2018mitigate} can be utilized to the dynamic calibration of cyber risks as a pre-screening method before the design of cyber insurance contracts.
The work \citep{chen2021dynamic} has investigated dynamic contract design to provide hints for asset owners and cyber risk managers to the cyber risk management of enterprise networks.

Apart from dynamic risk assessment, dynamic contract problems are themselves challenging.
It is straightforward to obtain a stochastic game problem when we consider the extension of the P-A problem to the dynamic setting.
One common approach to analyze and solve a dynamic P-A problem is to reformulate it to stochastic control problems and utilize dynamic programming.
This method is adopted, for example, in \citep{cvitanic2018dynamic,chen2021dynamic}.
The approach of \citep{zhang2021optimal} digs into the structures of some representative problem forms to find the optimal dynamic insurance plan. 
More generally, dynamic insurance design is closely related to dynamic mechanism design, which we refer to \citep{bergemann2019dynamic,pavan2014dynamic,zhang2022incentive} and the references therein.

The notion of cyber resilience naturally extends to its dynamic version.
The reason lies in the following.
Firstly, resilience itself involves a two-stage perspective towards cyber threats.
As we have discussed in Section \ref{sec:defense mechanisms:residual}, cyber resilience accounts for the mitigation of residual risk, which is captured by the cyber risk that remains after the effect of defense mechanisms.
Hence, from the perspective of cyber resilience, cyber threats are no longer static.
Secondly, one of the consequences of enhancing cyber resilience is the restoration of a system from the damage induced by cyber attacks.
While this recovery represents that the system survives the previous cyber threats, it also puts the system in front of future attacks.
Hence, users and insurers naturally commit consecutive efforts to either invest on system protection or update the insurance contract.

In conclusion, there is a need for future cyber insurance markets to consider dynamic insurance contracts to cope with the ever-changing cyber threats and to reach a reasonable level of cyber resilience.

\section{Regulations on Cyber Insurance}
\label{sec:regulations}
Since the rapid growth of the cyber insurance market only starts recently, cyber insurance is a fairly new product  which not all users are totally familiar with.
Regulations on cyber insurance can help standardize the market and restrict certain speculative behaviors of both the insurer and the insuree.

From the perspective of policies, voluntary participation in cyber insurance seems to be the right policy when an insurance market is established.
However, since many businesses and companies are service providers who may store private information of customers or take actions on behalf of the users, there is a need to set certain requirements on the enrollment of cyber insurance not only to prevent severe losses to the businesses but also to protect the privacy and rights of the customers.
For example, risky behaviors of service providers can be extremely harmful to the customers and users when their information or assets are the used as resources by the service providers to compete in the market.
Whether or not the service providers have the rights to take risks on behalf of the users is not only an issue related to making profits but also an issue concerning responsibility, duty, and law.
Two simple approaches can be considered to normalize insurance participation.
The first approach is making cyber insurance a mandate for service providers.
Since there is always a possibility for residual risks to occur, a mandatory participation in cyber insurance offsets the risks of the customers.
With the coverage promised by the insurance contract, even when the providers become victims of cyber attacks, they can make compensation for the users and customers instead of suspending their services.
Mandates on insurance are particularly important when the service providers do not possess a high level of risk-aversion.
The second approach is to bundle cyber insurance products with IT products such as operating systems and networking devices.
The fact that this policy targets products instead of individuals or businesses makes it more acceptable for users.
Besides, users can also gain knowledge about the coverage and uses of cyber insurance contract when they select IT products.

From the perspective of the cyber insurance market, regulations are necessary in completing the supply chain.
The fundamental market policies start from education of the existence of cyber threats and the resilience that cyber insurance can provide.
The awareness of the security risks serves as the incentives of the users in participating in insurance products.
The education can also involve a benign perception manipulation procedure to further help the customers in understand the fatality of various cyber attacks and the failures they can induce on networked systems.
Perhaps standards can be set up on the users' risk preferences before they become service providers.
Once the users obtain the awareness of the potential harms of cyber threats through the education, incentive mechanisms can be deployed more fluently, for the users are more likely to agree on sharing the risk with an insurer and with the other insurees.

Apart from the above regulatory aspects of cyber insurance, it is also necessary to design accountability so that certain penalties will be induced to the party who violates certain rules prescribed in the insurance contracts.
A class of actions that should be restricted but can be easily ignored are those that exhibits opportunism.
These actions can seem promising when the action takers do not fully present the risks associated with the advantageous outcomes.
In fact, risk-seeking can be interpreted as actions exhibiting opportunism, since risk-seekers tend to focus on the value of the outcome, often in its best possible scenario, instead of the averaged outcome.
In the context of cybersecurity, such actions include, for example, running a large networked system with an out-dated firewall to save expenses.
Therefore, these actions can, with high probabilities, generate far less profits than that in the ideal scenario.
Modern risk measures can be utilized to design the accountability for this type of actions, since they naturally balance between the value and the possibility of outcomes.

\begin{comment}
\begin{table}
\caption{Enter table caption here.\label{tab1-1}}{%
\begin{tabular}{@{}cccc@{}}
\toprule
Tap     &Relative   &Relative   &Relative mean\\
number  &power (dB) &delay (ns) &power (dB)\\
\midrule
3 &0$-9.0$  &68,900\footnotemark[1] &$-12.8$\\
4 &$-10.0$ &12,900\footnotemark[2] &$-10.0$\\
5 &$-15.0$ &17,100 &$-25.2$\\
\botrule
\end{tabular}}{\footnotetext[]{Source: Example for table source text.}
\footnotetext[1]{Example for a first table footnote. Example for a first table footnote. Example for a first table footnote. Example for a first table footnote.}
\footnotetext[2]{Example for a second table footnote.}}
\end{table}
\end{comment}

\backmatter

\bibliographystyle{plainnat}
\bibliography{wiley}%

%% file: sample1.bbl
\begin{thebibliography}{113}
\providecommand{\natexlab}[1]{#1}
\providecommand{\url}[1]{\texttt{#1}}
\expandafter\ifx\csname urlstyle\endcsname\relax
  \providecommand{\doi}[1]{doi: #1}\else
  \providecommand{\doi}{doi: \begingroup \urlstyle{rm}\Url}\fi

\bibitem[Aidan et~al.(2017)Aidan, Verma, and Awasthi]{aidan2017comprehensive}
Jagmeet~Singh Aidan, Harsh~Kumar Verma, and Lalit~Kumar Awasthi.
\newblock Comprehensive survey on petya ransomware attack.
\newblock In \emph{2017 International Conference on Next Generation Computing
  and Information Systems (ICNGCIS)}, pages 122--125. IEEE, 2017.

\bibitem[Al-Rabiaah(2018)]{al2018stuxnet}
Sumayah Al-Rabiaah.
\newblock The “stuxnet” virus of 2010 as an example of a “apt” and its
  “recent” variances.
\newblock In \emph{2018 21st Saudi Computer Society National Computer
  Conference (NCC)}, pages 1--5. IEEE, 2018.

\bibitem[Al-rimy et~al.(2018)Al-rimy, Maarof, and Shaid]{al2018ransomware}
Bander Ali~Saleh Al-rimy, Mohd~Aizaini Maarof, and Syed Zainudeen~Mohd Shaid.
\newblock Ransomware threat success factors, taxonomy, and countermeasures: A
  survey and research directions.
\newblock \emph{Computers \& Security}, 74:\penalty0 144--166, 2018.

\bibitem[Almashhadani et~al.(2019)Almashhadani, Kaiiali, Sezer, and
  O’Kane]{almashhadani2019multi}
Ahmad~O Almashhadani, Mustafa Kaiiali, Sakir Sezer, and Philip O’Kane.
\newblock A multi-classifier network-based crypto ransomware detection system:
  A case study of locky ransomware.
\newblock \emph{IEEE access}, 7:\penalty0 47053--47067, 2019.

\bibitem[Alpcan and Basar(2006)]{alpcan2006intrusion}
Tansu Alpcan and Tamer Basar.
\newblock An intrusion detection game with limited observations.
\newblock In \emph{12th Int. Symp. on Dynamic Games and Applications, Sophia
  Antipolis, France}, volume~26, 2006.

\bibitem[Alshamrani et~al.(2019)Alshamrani, Myneni, Chowdhary, and
  Huang]{alshamrani2019survey}
Adel Alshamrani, Sowmya Myneni, Ankur Chowdhary, and Dijiang Huang.
\newblock A survey on advanced persistent threats: Techniques, solutions,
  challenges, and research opportunities.
\newblock \emph{IEEE Communications Surveys \& Tutorials}, 21\penalty0
  (2):\penalty0 1851--1877, 2019.

\bibitem[An et~al.(2012)An, Shieh, Tambe, Yang, Baldwin, DiRenzo, Maule, and
  Meyer]{an2012protect}
Bo~An, Eric Shieh, Milind Tambe, Rong Yang, Craig Baldwin, Joseph DiRenzo, Ben
  Maule, and Garrett Meyer.
\newblock Protect--a deployed game theoretic system for strategic security
  allocation for the united states coast guard.
\newblock \emph{Ai Magazine}, 33\penalty0 (4):\penalty0 96--96, 2012.

\bibitem[Anderson and Mellor(2009)]{anderson2009risk}
Lisa~R Anderson and Jennifer~M Mellor.
\newblock Are risk preferences stable? comparing an experimental measure with a
  validated survey-based measure.
\newblock \emph{Journal of Risk and Uncertainty}, 39\penalty0 (2):\penalty0
  137--160, 2009.

\bibitem[Arrow(1971)]{arrow1971theory}
Kenneth~J Arrow.
\newblock The theory of risk aversion.
\newblock \emph{Essays in the theory of risk-bearing}, pages 90--120, 1971.

\bibitem[Artzner et~al.(1999)Artzner, Delbaen, Eber, and
  Heath]{artzner1999coherent}
Philippe Artzner, Freddy Delbaen, Jean-Marc Eber, and David Heath.
\newblock Coherent measures of risk.
\newblock \emph{Mathematical finance}, 9\penalty0 (3):\penalty0 203--228, 1999.

\bibitem[Aslan and Samet(2020)]{aslan2020comprehensive}
{\"O}mer~Aslan Aslan and Refik Samet.
\newblock A comprehensive review on malware detection approaches.
\newblock \emph{IEEE Access}, 8:\penalty0 6249--6271, 2020.

\bibitem[Badhwar(2021)]{badhwar2021dynamic}
Raj Badhwar.
\newblock Dynamic measurement of cyber risk.
\newblock In \emph{The CISO’s Next Frontier}, pages 327--334. Springer, 2021.

\bibitem[Barona and Anita(2017)]{barona2017survey}
R~Barona and EA~Mary Anita.
\newblock A survey on data breach challenges in cloud computing security:
  Issues and threats.
\newblock In \emph{2017 International conference on circuit, power and
  computing technologies (ICCPCT)}, pages 1--8. IEEE, 2017.

\bibitem[Ba{\c{s}}ar and Olsder(1998)]{bacsar1998dynamic}
Tamer Ba{\c{s}}ar and Geert~Jan Olsder.
\newblock \emph{Dynamic noncooperative game theory}.
\newblock SIAM, 1998.

\bibitem[Bergemann and V{\"a}lim{\"a}ki(2019)]{bergemann2019dynamic}
Dirk Bergemann and Juuso V{\"a}lim{\"a}ki.
\newblock Dynamic mechanism design: An introduction.
\newblock \emph{Journal of Economic Literature}, 57\penalty0 (2):\penalty0
  235--74, 2019.

\bibitem[Bertsimas and Takeda(2015)]{bertsimas2015optimizing}
Dimitris Bertsimas and Akiko Takeda.
\newblock Optimizing over coherent risk measures and non-convexities: a robust
  mixed integer optimization approach.
\newblock \emph{Computational Optimization and Applications}, 62\penalty0
  (3):\penalty0 613--639, 2015.

\bibitem[Bishop et~al.(2009)Bishop, Engle, Peisert, Whalen, and
  Gates]{bishop2009case}
Matt Bishop, Sophie Engle, Sean Peisert, Sean Whalen, and Carrie Gates.
\newblock Case studies of an insider framework.
\newblock In \emph{2009 42nd Hawaii International Conference on System
  Sciences}, pages 1--10. IEEE, 2009.

\bibitem[B{\"o}hme et~al.(2010)B{\"o}hme, Schwartz, et~al.]{bohme2010modeling}
Rainer B{\"o}hme, Galina Schwartz, et~al.
\newblock Modeling cyber-insurance: towards a unifying framework.
\newblock In \emph{WEIS}, 2010.

\bibitem[Braue(2022)]{RansomwarePrediction}
David Braue.
\newblock Global ransomware damage costs predicted to exceed $265$ billion by
  $2031$, 2022.
\newblock URL
  \url{https://cybersecurityventures.com/global-ransomware-damage-costs-predicted-to-reach-250-billion-usd-by-2031/}.

\bibitem[Casey et~al.(2016)Casey, Morales, Wright, Zhu, and
  Mishra]{casey2016compliance}
William Casey, Jose~Andre Morales, Evan Wright, Quanyan Zhu, and Bud Mishra.
\newblock Compliance signaling games: toward modeling the deterrence of insider
  threats.
\newblock \emph{Computational and Mathematical Organization Theory},
  22\penalty0 (3):\penalty0 318--349, 2016.

\bibitem[Chandran et~al.(2015)Chandran, Hrudya, and
  Poornachandran]{chandran2015efficient}
Saranya Chandran, P~Hrudya, and Prabaharan Poornachandran.
\newblock An efficient classification model for detecting advanced persistent
  threat.
\newblock In \emph{2015 international conference on advances in computing,
  communications and informatics (ICACCI)}, pages 2001--2009. IEEE, 2015.

\bibitem[Chen et~al.(2021)Chen, Zhu, and Ba{\c{s}}ar]{chen2021dynamic}
Juntao Chen, Quanyan Zhu, and Tamer Ba{\c{s}}ar.
\newblock Dynamic contract design for systemic cyber risk management of
  interdependent enterprise networks.
\newblock \emph{Dynamic Games and Applications}, 11\penalty0 (2):\penalty0
  294--325, 2021.

\bibitem[Chen and Bridges(2017)]{chen2017automated}
Qian Chen and Robert~A Bridges.
\newblock Automated behavioral analysis of malware: A case study of wannacry
  ransomware.
\newblock In \emph{2017 16th IEEE International Conference on Machine Learning
  and Applications (ICMLA)}, pages 454--460. IEEE, 2017.

\bibitem[Colwill(2009)]{colwill2009human}
Carl Colwill.
\newblock Human factors in information security: The insider threat--who can
  you trust these days?
\newblock \emph{Information security technical report}, 14\penalty0
  (4):\penalty0 186--196, 2009.

\bibitem[Crawford(1985)]{crawford1985dynamic}
Vincent~P Crawford.
\newblock Dynamic games and dynamic contract theory.
\newblock \emph{Journal of Conflict resolution}, 29\penalty0 (2):\penalty0
  195--224, 1985.

\bibitem[Cvitani{\'c} et~al.(2018)Cvitani{\'c}, Possama{\"\i}, and
  Touzi]{cvitanic2018dynamic}
Jak{\v{s}}a Cvitani{\'c}, Dylan Possama{\"\i}, and Nizar Touzi.
\newblock Dynamic programming approach to principal--agent problems.
\newblock \emph{Finance and Stochastics}, 22\penalty0 (1):\penalty0 1--37,
  2018.

\bibitem[Dempe(2002)]{dempe2002foundations}
Stephan Dempe.
\newblock \emph{Foundations of bilevel programming}.
\newblock Springer Science \& Business Media, 2002.

\bibitem[El-Kosairy and Azer(2018)]{el2018intrusion}
Ahmed El-Kosairy and Marianne~A Azer.
\newblock Intrusion and ransomware detection system.
\newblock In \emph{2018 1st International Conference on Computer Applications
  \& Information Security (ICCAIS)}, pages 1--7. IEEE, 2018.

\bibitem[Fang et~al.(2013)Fang, Jiang, and Tambe]{fang2013protecting}
Fei Fang, Albert~Xin Jiang, and Milind Tambe.
\newblock Protecting moving targets with multiple mobile resources.
\newblock \emph{Journal of Artificial Intelligence Research}, 48:\penalty0
  583--634, 2013.

\bibitem[Farwell and Rohozinski(2011)]{farwell2011stuxnet}
James~P Farwell and Rafal Rohozinski.
\newblock Stuxnet and the future of cyber war.
\newblock \emph{Survival}, 53\penalty0 (1):\penalty0 23--40, 2011.

\bibitem[Feng et~al.(2017)Feng, Zheng, Mohapatra, and
  Cansever]{feng2017stackelberg}
Xiaotao Feng, Zizhan Zheng, Prasant Mohapatra, and Derya Cansever.
\newblock A stackelberg game and markov modeling of moving target defense.
\newblock In \emph{International Conference on Decision and Game Theory for
  Security}, pages 315--335. Springer, 2017.

\bibitem[F{\"o}llmer and Schied(2016)]{follmer2016stochastic}
Hans F{\"o}llmer and Alexander Schied.
\newblock Stochastic finance.
\newblock In \emph{Stochastic Finance}. de Gruyter, 2016.

\bibitem[Garbis and Chapman(2021)]{garbis2021zero}
Jason Garbis and Jerry~W Chapman.
\newblock \emph{Zero Trust Security: An Enterprise Guide}.
\newblock Springer, 2021.

\bibitem[Ghafir et~al.(2018)Ghafir, Hammoudeh, Prenosil, Han, Hegarty, Rabie,
  and Aparicio-Navarro]{ghafir2018detection}
Ibrahim Ghafir, Mohammad Hammoudeh, Vaclav Prenosil, Liangxiu Han, Robert
  Hegarty, Khaled Rabie, and Francisco~J Aparicio-Navarro.
\newblock Detection of advanced persistent threat using machine-learning
  correlation analysis.
\newblock \emph{Future Generation Computer Systems}, 89:\penalty0 349--359,
  2018.

\bibitem[Grossman and Hart(1992)]{grossman1992analysis}
Sanford~J Grossman and Oliver~D Hart.
\newblock An analysis of the principal-agent problem.
\newblock In \emph{Foundations of insurance economics}, pages 302--340.
  Springer, 1992.

\bibitem[Gunson et~al.(2011)Gunson, Marshall, Morton, and Jack]{gunson2011user}
Nancie Gunson, Diarmid Marshall, Hazel Morton, and Mervyn Jack.
\newblock User perceptions of security and usability of single-factor and
  two-factor authentication in automated telephone banking.
\newblock \emph{Computers \& Security}, 30\penalty0 (4):\penalty0 208--220,
  2011.

\bibitem[Harsanyi(1967)]{harsanyi1967games}
John~C Harsanyi.
\newblock Games with incomplete information played by “bayesian” players,
  i--iii part i. the basic model.
\newblock \emph{Management science}, 14\penalty0 (3):\penalty0 159--182, 1967.

\bibitem[Hobbs(2021)]{hobbs2021colonial}
Allegra Hobbs.
\newblock The colonial pipeline hack: Exposing vulnerabilities in us
  cybersecurity.
\newblock In \emph{SAGE Business Cases}. SAGE Publications: SAGE Business Cases
  Originals, 2021.

\bibitem[Holmstr{\"o}m(1979)]{holmstrom1979moral}
Bengt Holmstr{\"o}m.
\newblock Moral hazard and observability.
\newblock \emph{The Bell journal of economics}, pages 74--91, 1979.

\bibitem[Huang and Zhu(2019{\natexlab{a}})]{huang2019adaptive}
Linan Huang and Quanyan Zhu.
\newblock Adaptive strategic cyber defense for advanced persistent threats in
  critical infrastructure networks.
\newblock \emph{ACM SIGMETRICS Performance Evaluation Review}, 46\penalty0
  (2):\penalty0 52--56, 2019{\natexlab{a}}.

\bibitem[Huang and Zhu(2019{\natexlab{b}})]{huang2019adaptivehoneypot}
Linan Huang and Quanyan Zhu.
\newblock Adaptive honeypot engagement through reinforcement learning of
  semi-markov decision processes.
\newblock In \emph{International conference on decision and game theory for
  security}, pages 196--216. Springer, 2019{\natexlab{b}}.

\bibitem[Huang and Zhu(2019{\natexlab{c}})]{huang2019dynamic}
Linan Huang and Quanyan Zhu.
\newblock Dynamic bayesian games for adversarial and defensive cyber deception.
\newblock In \emph{Autonomous cyber deception}, pages 75--97. Springer,
  2019{\natexlab{c}}.

\bibitem[Huang and Zhu(2020)]{huang2020dynamic}
Linan Huang and Quanyan Zhu.
\newblock A dynamic games approach to proactive defense strategies against
  advanced persistent threats in cyber-physical systems.
\newblock \emph{Computers \& Security}, 89:\penalty0 101660, 2020.

\bibitem[Huang and Zhu(2021)]{huang2021duplicity}
Linan Huang and Quanyan Zhu.
\newblock Duplicity games for deception design with an application to insider
  threat mitigation.
\newblock \emph{IEEE Transactions on Information Forensics and Security},
  16:\penalty0 4843--4856, 2021.

\bibitem[Huang and Zhu(2019{\natexlab{d}})]{huang2019deceptive}
Yunhan Huang and Quanyan Zhu.
\newblock Deceptive reinforcement learning under adversarial manipulations on
  cost signals.
\newblock In \emph{International Conference on Decision and Game Theory for
  Security}, pages 217--237. Springer, 2019{\natexlab{d}}.

\bibitem[Huang et~al.(2020)Huang, Chen, Huang, and Zhu]{yhuang2020dynamic}
Yunhan Huang, Juntao Chen, Linan Huang, and Quanyan Zhu.
\newblock Dynamic games for secure and resilient control system design.
\newblock \emph{National Science Review}, 7\penalty0 (7):\penalty0 1125--1141,
  2020.

\bibitem[Huang et~al.(2022)Huang, Huang, and Zhu]{huang2022reinforcement}
Yunhan Huang, Linan Huang, and Quanyan Zhu.
\newblock Reinforcement learning for feedback-enabled cyber resilience.
\newblock \emph{Annual Reviews in Control}, 2022.

\bibitem[Hunker and Probst(2011)]{hunker2011insiders}
Jeffrey Hunker and Christian~W Probst.
\newblock Insiders and insider threats-an overview of definitions and
  mitigation techniques.
\newblock \emph{J. Wirel. Mob. Networks Ubiquitous Comput. Dependable Appl.},
  2\penalty0 (1):\penalty0 4--27, 2011.

\bibitem[Hutchins et~al.(2011)Hutchins, Cloppert, Amin,
  et~al.]{hutchins2011intelligence}
Eric~M Hutchins, Michael~J Cloppert, Rohan~M Amin, et~al.
\newblock Intelligence-driven computer network defense informed by analysis of
  adversary campaigns and intrusion kill chains.
\newblock \emph{Leading Issues in Information Warfare \& Security Research},
  1\penalty0 (1):\penalty0 80, 2011.

\bibitem[Jain et~al.(2010)Jain, Tsai, Pita, Kiekintveld, Rathi, Tambe, and
  Ord{\'o}nez]{jain2010software}
Manish Jain, Jason Tsai, James Pita, Christopher Kiekintveld, Shyamsunder
  Rathi, Milind Tambe, and Fernando Ord{\'o}nez.
\newblock Software assistants for randomized patrol planning for the lax
  airport police and the federal air marshal service.
\newblock \emph{Interfaces}, 40\penalty0 (4):\penalty0 267--290, 2010.

\bibitem[Joshi et~al.(2020)Joshi, Aliaga, and Insua]{joshi2020insider}
Chaitanya Joshi, Jesus~Rios Aliaga, and David~Rios Insua.
\newblock Insider threat modeling: An adversarial risk analysis approach.
\newblock \emph{IEEE Transactions on Information Forensics and Security},
  16:\penalty0 1131--1142, 2020.

\bibitem[Khalili et~al.(2018)Khalili, Naghizadeh, and
  Liu]{khalili2018designing}
Mohammad~Mahdi Khalili, Parinaz Naghizadeh, and Mingyan Liu.
\newblock Designing cyber insurance policies: The role of pre-screening and
  security interdependence.
\newblock \emph{IEEE Transactions on Information Forensics and Security},
  13\penalty0 (9):\penalty0 2226--2239, 2018.

\bibitem[Kurt et~al.(2018)Kurt, Ogundijo, Li, and Wang]{kurt2018online}
Mehmet~Necip Kurt, Oyetunji Ogundijo, Chong Li, and Xiaodong Wang.
\newblock Online cyber-attack detection in smart grid: A reinforcement learning
  approach.
\newblock \emph{IEEE Transactions on Smart Grid}, 10\penalty0 (5):\penalty0
  5174--5185, 2018.

\bibitem[Kusuoka(2001)]{kusuoka2001law}
Shigeo Kusuoka.
\newblock On law invariant coherent risk measures.
\newblock In \emph{Advances in mathematical economics}, pages 83--95. Springer,
  2001.

\bibitem[Levy(1992)]{levy1992stochastic}
Haim Levy.
\newblock Stochastic dominance and expected utility: Survey and analysis.
\newblock \emph{Management science}, 38\penalty0 (4):\penalty0 555--593, 1992.

\bibitem[Liao et~al.(2016)Liao, Zhao, Doup{\'e}, and Ahn]{liao2016behind}
Kevin Liao, Ziming Zhao, Adam Doup{\'e}, and Gail-Joon Ahn.
\newblock Behind closed doors: measurement and analysis of cryptolocker ransoms
  in bitcoin.
\newblock In \emph{2016 APWG symposium on electronic crime research (eCrime)},
  pages 1--13. IEEE, 2016.

\bibitem[Liu et~al.(2018{\natexlab{a}})Liu, De~Vel, Han, Zhang, and
  Xiang]{liu2018detecting}
Liu Liu, Olivier De~Vel, Qing-Long Han, Jun Zhang, and Yang Xiang.
\newblock Detecting and preventing cyber insider threats: A survey.
\newblock \emph{IEEE Communications Surveys \& Tutorials}, 20\penalty0
  (2):\penalty0 1397--1417, 2018{\natexlab{a}}.

\bibitem[Liu et~al.(2018{\natexlab{b}})Liu, Han, Wang, and
  Zhou]{liu2018understanding}
Liyuan Liu, Meng Han, Yan Wang, and Yiyun Zhou.
\newblock Understanding data breach: A visualization aspect.
\newblock In \emph{International Conference on Wireless Algorithms, Systems,
  and Applications}, pages 883--892. Springer, 2018{\natexlab{b}}.

\bibitem[Liu and Zhu(2020)]{liu2020robust}
Shutian Liu and Quanyan Zhu.
\newblock Robust and stochastic optimization with a hybrid coherent risk
  measure with an application to supervised learning.
\newblock \emph{IEEE Control Systems Letters}, 5\penalty0 (3):\penalty0
  965--970, 2020.

\bibitem[Liu and Zhu(2022{\natexlab{a}})]{liu2022mitigating}
Shutian Liu and Quanyan Zhu.
\newblock Mitigating moral hazard in cyber insurance using risk preference
  design.
\newblock \emph{arXiv preprint arXiv:2203.12001}, 2022{\natexlab{a}}.

\bibitem[Liu and Zhu(2022{\natexlab{b}})]{liu2022role}
Shutian Liu and Quanyan Zhu.
\newblock On the role of risk perceptions in cyber insurance contracts.
\newblock \emph{arXiv preprint arXiv:2210.15010}, 2022{\natexlab{b}}.

\bibitem[Marotta et~al.(2017)Marotta, Martinelli, Nanni, Orlando, and
  Yautsiukhin]{marotta2017cyber}
Angelica Marotta, Fabio Martinelli, Stefano Nanni, Albina Orlando, and Artsiom
  Yautsiukhin.
\newblock Cyber-insurance survey.
\newblock \emph{Computer Science Review}, 24:\penalty0 35--61, 2017.

\bibitem[Mc~Carthy et~al.(2016)Mc~Carthy, Sinha, Tambe, and
  Manadhata]{mc2016data}
Sara~Marie Mc~Carthy, Arunesh Sinha, Milind Tambe, and Pratyusa Manadhata.
\newblock Data exfiltration detection and prevention: Virtually distributed
  pomdps for practically safer networks.
\newblock In \emph{International Conference on Decision and Game Theory for
  Security}, pages 39--61. Springer, 2016.

\bibitem[Miehling et~al.(2018)Miehling, Rasouli, and
  Teneketzis]{miehling2018pomdp}
Erik Miehling, Mohammad Rasouli, and Demosthenis Teneketzis.
\newblock A pomdp approach to the dynamic defense of large-scale cyber
  networks.
\newblock \emph{IEEE Transactions on Information Forensics and Security},
  13\penalty0 (10):\penalty0 2490--2505, 2018.

\bibitem[Noorani and Baras(2022)]{noorani2022embracing}
Erfaun Noorani and John~S Baras.
\newblock Embracing risk in reinforcement learning: The connection between
  risk-sensitive exponential and distributionally robust criteria.
\newblock In \emph{2022 American Control Conference (ACC)}, pages 2703--2708.
  IEEE, 2022.

\bibitem[Ometov et~al.(2018)Ometov, Bezzateev, M{\"a}kitalo, Andreev, Mikkonen,
  and Koucheryavy]{ometov2018multi}
Aleksandr Ometov, Sergey Bezzateev, Niko M{\"a}kitalo, Sergey Andreev, Tommi
  Mikkonen, and Yevgeni Koucheryavy.
\newblock Multi-factor authentication: A survey.
\newblock \emph{Cryptography}, 2\penalty0 (1):\penalty0 1, 2018.

\bibitem[Papastergiou and Polemi(2018)]{papastergiou2018mitigate}
Spyridon Papastergiou and Nineta Polemi.
\newblock Mitigate: A dynamic supply chain cyber risk assessment methodology.
\newblock In \emph{Smart Trends in Systems, Security and Sustainability}, pages
  1--9. Springer, 2018.

\bibitem[Pauly(1968)]{pauly1968economics}
Mark~V Pauly.
\newblock The economics of moral hazard: comment.
\newblock \emph{The american economic review}, 58\penalty0 (3):\penalty0
  531--537, 1968.

\bibitem[Pavan et~al.(2014)Pavan, Segal, and Toikka]{pavan2014dynamic}
Alessandro Pavan, Ilya Segal, and Juuso Toikka.
\newblock Dynamic mechanism design: A myersonian approach.
\newblock \emph{Econometrica}, 82\penalty0 (2):\penalty0 601--653, 2014.

\bibitem[Pawlick and Zhu(2017)]{pawlick2017strategic}
Jeffrey Pawlick and Quanyan Zhu.
\newblock Strategic trust in cloud-enabled cyber-physical systems with an
  application to glucose control.
\newblock \emph{IEEE Transactions on Information Forensics and Security},
  12\penalty0 (12):\penalty0 2906--2919, 2017.

\bibitem[Pawlick and Zhu(2021)]{pawlick2021game}
Jeffrey Pawlick and Quanyan Zhu.
\newblock \emph{Game Theory for Cyber Deception}.
\newblock Springer, 2021.

\bibitem[Pawlick et~al.(2015)Pawlick, Farhang, and Zhu]{pawlick2015flip}
Jeffrey Pawlick, Sadegh Farhang, and Quanyan Zhu.
\newblock Flip the cloud: Cyber-physical signaling games in the presence of
  advanced persistent threats.
\newblock In \emph{International Conference on Decision and Game Theory for
  Security}, pages 289--308. Springer, 2015.

\bibitem[Pawlick et~al.(2019)Pawlick, Colbert, and Zhu]{pawlick2019game}
Jeffrey Pawlick, Edward Colbert, and Quanyan Zhu.
\newblock A game-theoretic taxonomy and survey of defensive deception for
  cybersecurity and privacy.
\newblock \emph{ACM Computing Surveys (CSUR)}, 52\penalty0 (4):\penalty0 1--28,
  2019.

\bibitem[Pfleeger et~al.(2009)Pfleeger, Predd, Hunker, and
  Bulford]{pfleeger2009insiders}
Shari~Lawrence Pfleeger, Joel~B Predd, Jeffrey Hunker, and Carla Bulford.
\newblock Insiders behaving badly: Addressing bad actors and their actions.
\newblock \emph{IEEE transactions on information forensics and security},
  5\penalty0 (1):\penalty0 169--179, 2009.

\bibitem[Pflug and Pichler(2014)]{pflug2014multistage}
Georg~Ch Pflug and Alois Pichler.
\newblock \emph{Multistage stochastic optimization}, volume 1104.
\newblock Springer, 2014.

\bibitem[Pita et~al.(2008)Pita, Jain, Marecki, Ord{\'o}{\~n}ez, Portway, Tambe,
  Western, Paruchuri, and Kraus]{pita2008deployed}
James Pita, Manish Jain, Janusz Marecki, Fernando Ord{\'o}{\~n}ez, Christopher
  Portway, Milind Tambe, Craig Western, Praveen Paruchuri, and Sarit Kraus.
\newblock Deployed armor protection: the application of a game theoretic model
  for security at the los angeles international airport.
\newblock In \emph{Proceedings of the 7th international joint conference on
  Autonomous agents and multiagent systems: industrial track}, pages 125--132,
  2008.

\bibitem[Pratt(1978)]{pratt1978risk}
John~W Pratt.
\newblock Risk aversion in the small and in the large.
\newblock In \emph{Uncertainty in economics}, pages 59--79. Elsevier, 1978.

\bibitem[Rahimian and Mehrotra(2019)]{rahimian2019distributionally}
Hamed Rahimian and Sanjay Mehrotra.
\newblock Distributionally robust optimization: A review.
\newblock \emph{arXiv preprint arXiv:1908.05659}, 2019.

\bibitem[Rass and Zhu(2016)]{rass2016gadapt}
Stefan Rass and Quanyan Zhu.
\newblock Gadapt: a sequential game-theoretic framework for designing
  defense-in-depth strategies against advanced persistent threats.
\newblock In \emph{International conference on decision and game theory for
  security}, pages 314--326. Springer, 2016.

\bibitem[Rockafellar et~al.(2000)Rockafellar, Uryasev,
  et~al.]{rockafellar2000optimization}
R~Tyrrell Rockafellar, Stanislav Uryasev, et~al.
\newblock Optimization of conditional value-at-risk.
\newblock \emph{Journal of risk}, 2:\penalty0 21--42, 2000.

\bibitem[Rose et~al.(2020)Rose, Borchert, Mitchell, and Connelly]{rose2020zero}
Scott Rose, Oliver Borchert, Stu Mitchell, and Sean Connelly.
\newblock Zero trust architecture.
\newblock Technical report, National Institute of Standards and Technology,
  2020.

\bibitem[Ruszczy{\'n}ski and Shapiro(2006)]{ruszczynski2006optimization}
Andrzej Ruszczy{\'n}ski and Alexander Shapiro.
\newblock Optimization of convex risk functions.
\newblock \emph{Mathematics of operations research}, 31\penalty0 (3):\penalty0
  433--452, 2006.

\bibitem[Salem et~al.(2008)Salem, Hershkop, and Stolfo]{salem2008survey}
Malek~Ben Salem, Shlomo Hershkop, and Salvatore~J Stolfo.
\newblock A survey of insider attack detection research.
\newblock \emph{Insider Attack and Cyber Security}, pages 69--90, 2008.

\bibitem[Sallhammar et~al.(2006)Sallhammar, Helvik, and
  Knapskog]{sallhammar2006stochastic}
Karin Sallhammar, Bjarne~E Helvik, and Svein~J Knapskog.
\newblock On stochastic modeling for integrated security and dependability
  evaluation.
\newblock \emph{J. Networks}, 1\penalty0 (5):\penalty0 31--42, 2006.

\bibitem[Sarabi et~al.(2016)Sarabi, Naghizadeh, Liu, and Liu]{sarabi2016risky}
Armin Sarabi, Parinaz Naghizadeh, Yang Liu, and Mingyan Liu.
\newblock Risky business: Fine-grained data breach prediction using business
  profiles.
\newblock \emph{Journal of Cybersecurity}, 2\penalty0 (1):\penalty0 15--28,
  2016.

\bibitem[Sarraute et~al.(2012)Sarraute, Buffet, and
  Hoffmann]{sarraute2012pomdps}
Carlos Sarraute, Olivier Buffet, and J{\"o}rg Hoffmann.
\newblock Pomdps make better hackers: Accounting for uncertainty in penetration
  testing.
\newblock In \emph{Twenty-Sixth AAAI Conference on Artificial Intelligence},
  2012.

\bibitem[Schildberg-H{\"o}risch(2018)]{schildberg2018risk}
Hannah Schildberg-H{\"o}risch.
\newblock Are risk preferences stable?
\newblock \emph{Journal of Economic Perspectives}, 32\penalty0 (2):\penalty0
  135--54, 2018.

\bibitem[Shapiro et~al.(2021)Shapiro, Dentcheva, and
  Ruszczynski]{shapiro2021lectures}
Alexander Shapiro, Darinka Dentcheva, and Andrzej Ruszczynski.
\newblock \emph{Lectures on stochastic programming: modeling and theory}.
\newblock SIAM, 2021.

\bibitem[Siddiqui et~al.(2016)Siddiqui, Khan, Ferens, and
  Kinsner]{siddiqui2016detecting}
Sana Siddiqui, Muhammad~Salman Khan, Ken Ferens, and Witold Kinsner.
\newblock Detecting advanced persistent threats using fractal dimension based
  machine learning classification.
\newblock In \emph{Proceedings of the 2016 ACM on international workshop on
  security and privacy analytics}, pages 64--69, 2016.

\bibitem[Silowash et~al.(2012)Silowash, Cappelli, Moore, Trzeciak, Shimeall,
  and Flynn]{silowash2012common}
George~J Silowash, Dawn~M Cappelli, Andrew~P Moore, Randall~F Trzeciak, Timothy
  Shimeall, and Lori Flynn.
\newblock Common sense guide to mitigating insider threats.
\newblock 2012.

\bibitem[Sinha et~al.(2018)Sinha, Fang, An, Kiekintveld, and
  Tambe]{sinha2018stackelberg}
Arunesh Sinha, Fei Fang, Bo~An, Christopher Kiekintveld, and Milind Tambe.
\newblock Stackelberg security games: Looking beyond a decade of success.
\newblock IJCAI, 2018.

\bibitem[Stackelberg et~al.(1952)]{stackelberg1952theory}
Heinrich~von Stackelberg et~al.
\newblock Theory of the market economy.
\newblock 1952.

\bibitem[Stafford(2020)]{stafford2020zero}
VA~Stafford.
\newblock Zero trust architecture.
\newblock \emph{NIST Special Publication}, 800:\penalty0 207, 2020.

\bibitem[Stole(2001)]{stole2001lectures}
Lars Stole.
\newblock Lectures on the theory of contracts and organizations.
\newblock \emph{Unpublished monograph}, 2001.

\bibitem[Tversky and Kahneman(1992)]{tversky1992advances}
Amos Tversky and Daniel Kahneman.
\newblock Advances in prospect theory: Cumulative representation of
  uncertainty.
\newblock \emph{Journal of Risk and uncertainty}, 5\penalty0 (4):\penalty0
  297--323, 1992.

\bibitem[Von~Neumann and Morgenstern(2007)]{von2007theory}
John Von~Neumann and Oskar Morgenstern.
\newblock Theory of games and economic behavior.
\newblock In \emph{Theory of games and economic behavior}. Princeton university
  press, 2007.

\bibitem[Wan et~al.(2018)Wan, Chang, Chen, and Wang]{wan2018feature}
Yu-Lun Wan, Jen-Chun Chang, Rong-Jaye Chen, and Shiuh-Jeng Wang.
\newblock Feature-selection-based ransomware detection with machine learning of
  data analysis.
\newblock In \emph{2018 3rd international conference on computer and
  communication systems (ICCCS)}, pages 85--88. IEEE, 2018.

\bibitem[Zhang and Zhu(2019)]{zhang2019mathtt}
Rui Zhang and Quanyan Zhu.
\newblock $\mathtt{FlipIn}$ : A game-theoretic cyber insurance framework for
  incentive-compatible cyber risk management of internet of things.
\newblock \emph{IEEE Transactions on Information Forensics and Security},
  15:\penalty0 2026--2041, 2019.

\bibitem[Zhang and Zhu(2021)]{zhang2021optimal}
Rui Zhang and Quanyan Zhu.
\newblock Optimal cyber-insurance contract design for dynamic risk management
  and mitigation.
\newblock \emph{IEEE Transactions on Computational Social Systems}, 2021.

\bibitem[Zhang et~al.(2017)Zhang, Zhu, and Hayel]{zhang2017bi}
Rui Zhang, Quanyan Zhu, and Yezekael Hayel.
\newblock A bi-level game approach to attack-aware cyber insurance of computer
  networks.
\newblock \emph{IEEE Journal on Selected Areas in Communications}, 35\penalty0
  (3):\penalty0 779--794, 2017.

\bibitem[Zhang and Zhu(2022)]{zhang2022incentive}
Tao Zhang and Quanyan Zhu.
\newblock On incentive compatibility in dynamic mechanism design with exit
  option in a markovian environment.
\newblock \emph{Dynamic Games and Applications}, 12\penalty0 (2):\penalty0
  701--745, 2022.

\bibitem[Zhao et~al.(2021)Zhao, Ge, and Zhu]{zhao2021combating}
Yuhan Zhao, Yunfei Ge, and Quanyan Zhu.
\newblock Combating ransomware in internet of things: A games-in-games approach
  for cross-layer cyber defense and security investment.
\newblock In \emph{International Conference on Decision and Game Theory for
  Security}, pages 208--228. Springer, 2021.

\bibitem[Zhu(2019)]{zhu2019game}
Quanyan Zhu.
\newblock Game theory for cyber deception: a tutorial.
\newblock In \emph{Proceedings of the 6th Annual Symposium on Hot Topics in the
  Science of Security}, pages 1--3, 2019.

\bibitem[Zhu and Ba{\c{s}}ar(2009)]{zhu2009dynamic}
Quanyan Zhu and Tamer Ba{\c{s}}ar.
\newblock Dynamic policy-based ids configuration.
\newblock In \emph{Proceedings of the 48h IEEE Conference on Decision and
  Control (CDC) held jointly with 2009 28th Chinese Control Conference}, pages
  8600--8605. IEEE, 2009.

\bibitem[Zhu and Ba{\c{s}}ar(2013)]{zhu2013game}
Quanyan Zhu and Tamer Ba{\c{s}}ar.
\newblock Game-theoretic approach to feedback-driven multi-stage moving target
  defense.
\newblock In \emph{International conference on decision and game theory for
  security}, pages 246--263. Springer, 2013.

\bibitem[Zhu and Rass(2018{\natexlab{a}})]{zhu2018game}
Quanyan Zhu and Stefan Rass.
\newblock Game theory meets network security: A tutorial.
\newblock In \emph{Proceedings of the 2018 ACM SIGSAC Conference on Computer
  and Communications Security}, pages 2163--2165, 2018{\natexlab{a}}.

\bibitem[Zhu and Rass(2018{\natexlab{b}})]{zhu2018multi}
Quanyan Zhu and Stefan Rass.
\newblock On multi-phase and multi-stage game-theoretic modeling of advanced
  persistent threats.
\newblock \emph{IEEE Access}, 6:\penalty0 13958--13971, 2018{\natexlab{b}}.

\bibitem[Zhu and Xu(2020)]{zhu2020cross}
Quanyan Zhu and Zhiheng Xu.
\newblock \emph{Cross-Layer Design for Secure and Resilient Cyber-Physical
  Systems}.
\newblock Springer, 2020.

\bibitem[Zhu et~al.(2009)Zhu, Fung, Boutaba, and Basar]{zhu2009game}
Quanyan Zhu, Carol Fung, Raouf Boutaba, and Tamer Basar.
\newblock A game-theoretical approach to incentive design in collaborative
  intrusion detection networks.
\newblock In \emph{2009 International Conference on Game Theory for Networks},
  pages 384--392. IEEE, 2009.

\bibitem[Zhu et~al.(2010{\natexlab{a}})Zhu, Li, Han, and
  Ba{\c{s}}ar]{zhu2010stochastic}
Quanyan Zhu, Husheng Li, Zhu Han, and Tamer Ba{\c{s}}ar.
\newblock A stochastic game model for jamming in multi-channel cognitive radio
  systems.
\newblock In \emph{2010 IEEE International Conference on Communications}, pages
  1--6. IEEE, 2010{\natexlab{a}}.

\bibitem[Zhu et~al.(2010{\natexlab{b}})Zhu, Tembine, and
  Ba{\c{s}}ar]{zhu2010heterogeneous}
Quanyan Zhu, Hamidou Tembine, and Tamer Ba{\c{s}}ar.
\newblock Heterogeneous learning in zero-sum stochastic games with incomplete
  information.
\newblock In \emph{49th IEEE conference on decision and control (CDC)}, pages
  219--224. IEEE, 2010{\natexlab{b}}.

\bibitem[Zhu et~al.(2012{\natexlab{a}})Zhu, Fung, Boutaba, and
  Basar]{zhu2012guidex}
Quanyan Zhu, Carol Fung, Raouf Boutaba, and Tamer Basar.
\newblock Guidex: A game-theoretic incentive-based mechanism for intrusion
  detection networks.
\newblock \emph{IEEE Journal on Selected Areas in Communications}, 30\penalty0
  (11):\penalty0 2220--2230, 2012{\natexlab{a}}.

\bibitem[Zhu et~al.(2012{\natexlab{b}})Zhu, Tembine, and
  Ba{\c{s}}ar]{zhu2012hybrid}
Quanyan Zhu, Hamidou Tembine, and Tamer Ba{\c{s}}ar.
\newblock Hybrid learning in stochastic games and its application in network
  security.
\newblock \emph{Reinforcement Learning and Approximate Dynamic Programming for
  Feedback Control}, pages 303--329, 2012{\natexlab{b}}.

\end{thebibliography}
